\documentclass[%
 reprint,
superscriptaddress,
 amsmath,amssymb,
 aps,
prl,
]{revtex4-1}
        
	\usepackage{feynmf}
	\usepackage{graphics}
	\usepackage{epsfig}
	\usepackage{graphicx}
	\usepackage{color}
	\usepackage{comment}
	\usepackage{ulem}
	\usepackage{tabularx}
	\usepackage{multirow}
	\usepackage{multibib}
    \usepackage{ulem}
    \usepackage{xcolor}
 \usepackage[scr=boondoxo]{mathalpha}

\usepackage{xspace}

\newcommand{\rucl}{\mbox{$\alpha$-RuCl\textsubscript{3}}\xspace}

\newcolumntype{C}{>{\centering\arraybackslash}X}
\begin{document}
	\unitlength = 1mm

\title{Theory of Intrinsic Phonon Thermal Hall Effect in \rucl}

\author{Ramesh Dhakal}
\affiliation{Department of Physics, Wake Forest University, Winston-Salem, North Carolina 27109, USA}
\affiliation{Center for Functional Materials, Wake Forest University, Winston-Salem, North Carolina 27109, USA}

\author{David A. S. Kaib}
\affiliation{Institute of Theoretical Physics, Goethe University Frankfurt, Max-von-Laue-Straße 1, 60438 Frankfurt am Main, Germany}

\author{Kate Choi}
\affiliation{Department of Physics, Wake Forest University, Winston-Salem, North Carolina 27109, USA}

\author{Sananda Biswas}
\affiliation{Institute of Theoretical Physics, Goethe University Frankfurt, Max-von-Laue-Straße 1, 60438 Frankfurt am Main, Germany}

\author{Roser Valent\'i}
\affiliation{Institute of Theoretical Physics, Goethe University Frankfurt, Max-von-Laue-Straße 1, 60438 Frankfurt am Main, Germany}

\author{Stephen M. Winter}
\email{winters@wfu.edu}
\affiliation{Department of Physics, Wake Forest University, Winston-Salem, North Carolina 27109, USA}
\affiliation{Center for Functional Materials, Wake Forest University, Winston-Salem, North Carolina 27109, USA}
\date{\today}

\begin{abstract}

The observation of a sizable thermal Hall effect in the spin–orbit–coupled magnet \rucl, as well as in other magnetic insulators, has prompted considerable debate about the origin of this phenomenon. At the heart of this discussion lies the question of which heat carriers contribute to the thermal Hall effect, potentially involving both magnetic excitations and phonons, and which mechanism is responsible for generating a finite thermal Hall conductivity. 
 In this work, we employ a recently developed first-principles framework capable of treating general spin–phonon couplings in systems with strong spin–orbit interaction to investigate \rucl. We demonstrate that spin–orbit coupling significantly enriches the form of these couplings, and endows them with chirality conducive to generating finite phonon Berry curvature. We show that this leads to a phonon thermal Hall effect that qualitatively reproduces the experimentally measured field dependence of $\kappa_{xy}$ in \rucl without invoking the presence of a field-induced spin-liquid state. These results provide compelling evidence toward resolving the microscopic origin of the field-induced behavior in \rucl.

\end{abstract}

\maketitle

{\bf INTRODUCTION}

\vspace{0.2cm}

Reports of a  large thermal Hall effect (THE) in \mbox{\rucl} \cite{kasahara2018unusual,kasahara2018majorana,yokoi2021half,imamura2024majorana,hentrich2019large,lefranccois2022evidence,bruin2022robustness,czajka2023planar,zhang2024stacking}, and an increasing range of other magnetic insulators \cite{zhang2024thermal,ideue2017giant,zhang2021anomalous,xu2023thermal,xu2024thermal,
kim2024thermal,nawwar2025large}, has intensified discussion of the heat carriers contributing to the THE, which may include magnetic excitations (e.g.~spinons \cite{kitaev2006anyons,katsura2010theory,nasu2017thermal,kasahara2018unusual,kasahara2018majorana,yokoi2021half,imamura2024majorana} 
and topological magnons \cite{mook2014magnon,mcclarty2018topological,zhang2021topological,chern2021sign,li2022thermal,czajka2023planar,czajka2023planar}) as well as phonons. Various mechanisms have been proposed to generate a finite phonon thermal Hall conductivity $\kappa_{xy}$. These include {\it extrinsic} skew-scattering from magnetic impurities \cite{mori2014origin,guo2021extrinsic,sun2022large,guo2022resonant} as well as {\it intrinsic} effects such as direct coupling of charged ions to external magnetic fields \cite{saito2019berry,flebus2022charged,flebus2023phonon},
 the formation of topological magnon-polarons \cite{zhang2019thermal,thingstad2019chiral,
huang2021topological,bao2023direct}, spin-phonon scattering from bulk magnetic excitations \cite{ye2020phonon,ye2021phonon,mangeolle2022phonon,mangeolle2022thermal,zhang2021phonon,singh2024phonon}, and anomalous phonon velocities induced by phenomenological two-phonon ``Raman'' spin-phonon interactions \cite{sheng2006theory,wang2009phonon,zhang2010topological}. The varying phenomenology of different materials points to possible contributions from different heat carriers and mechanisms. For \rucl, initial reports of $\kappa_{xy}/T$ on the order of the half-quantum of thermal conductivity at intermediate fields led to speculation about possible contributions of Majorana spinon edge currents. However, subsequent studies also raised the possibility of non-negligible contributions from phonons \cite{lefranccois2022evidence,li2023magnons}, which are compatible with experimental evidence for strong magnetoelastic coupling \cite{hentrich2018unusual,gass2020FieldinducedTransitionsKitaev,schonemann2020ThermalMagnetoelasticPropertiesb,
kocsis2022MagnetoelasticCouplingAnisotropy}. 
 Part of the continuing ambiguity stems from the lack of a detailed understanding of the form and magnitude of the spin--phonon couplings and their consequences for thermal transport. Given the importance of the THE in identifying topologically ordered ground states \cite{kane1997quantized,kitaev2006anyons,banerjee2018observation}, a complete understanding of phononic contributions is essential.

While the theory of spin-lattice coupling is, in principle, well-developed \cite{
orbach1961spin,
valenti1999novel},
recent advances in first-principles methods \cite{
ren2024adiabatic,dhakal2024mn} now make it possible to address these questions in material-specific studies. In this work, we 
provide a comprehensive analysis of spin-phonon couplings in \mbox{\rucl} via 
first-principles calculations \cite{dhakal2024mn} and use them to assess the phononic contribution to the THE arising from {\it intrinsic} phonon Hall viscosity \cite{barkeshli2012dissipationless,ye2020phonon,ye2021phonon}. We first review the origin of the phonon Hall viscosity, and identify the relevant spin-phonon couplings. We then present the computed couplings, focusing on the acoustic phonon modes relevant to low-temperature transport. Finally, we evaluate the longitudinal and transverse phonon thermal conductivities as a function of magnetic field. We ultimately find that the intrinsic phonon Hall effect reproduces all aspects of the experimental low-temperature $\kappa_{xy}$.

In the low-energy description of lattice phonons, the phonon Hall viscosity is the leading term capturing time-reversal symmetry breaking. In an effective phonon Hamiltonian $\mathcal{H}_{\rm eff}$ such a term arises upon integrating out the electronic degrees of freedom and acts as an emergent gauge field on the phonons due to electron-phonon coupling \cite{mead1992geometric,barkeshli2012dissipationless}. Including this effect, $\mathcal{H}_{\rm eff}$ takes the form:
\begin{align}\label{eq:Heffph}
\mathcal{H}_{\rm eff} = & \ \mathcal{H}_0 -\frac{1}{2} \sum_q \left( \mathbf{u}_{q}^\dagger \mathbb{N}_q \mathbb{M}_q^{-1} \mathbf{p}_q - \mathbf{p}_q^\dagger \mathbb{M}_q^{-1} \mathbb{N}_q \mathbf{u}_q \right)
\end{align}
where $\mathcal{H}_0$ is the unperturbed phonon Hamiltonian,  $\mathbf{u}_q^\dagger = [u_{q,1}^\dagger \ u_{q,2}^\dagger \ ...]$ and $\mathbf{p}_q^\dagger = [p_{q,1}^\dagger \ p_{q,2}^\dagger \ ...]$ are the phonon displacement and momentum operators for different phonon bands, and $\mathbb{M}_q$ is the diagonal mass tensor. $\mathbb{N}_q$ is an antisymmetric matrix, whose long-wavelength limit reduces to the phonon Hall viscosity tensor discussed in Ref.~\cite{barkeshli2012dissipationless,ye2020phonon,ye2021phonon}.

To account for the effects of electron-phonon coupling on phonon dynamics in magnetic insulators, we start by considering the linear coupling between the electronic and lattice degrees of freedom via:\begin{align}\label{eq:H_el_ph}
\mathcal{H}_{\rm el-ph} = \sum_{q\nu} \mathcal{O}_{q\nu} u_{q\nu}
\end{align}
where $\mathcal{O}_{q\nu}$ is a generic operator acting on the electronic system, and $u_{q\nu}$
is the phonon displacement operator for momentum $q$ and band index $\nu$. After integrating out the electronic degrees of freedom, the lowest-order contribution to the phonon self-energy is \cite{ye2020phonon,ye2021phonon}:
\begin{align}\label{eq:phSE}
\Pi_{q}^{\nu\nu^\prime}(i\omega_n) = -\frac{1}{2\hbar}\int_{-\hbar\beta}^{\hbar\beta} d\tau\,   e^{i\omega_n \tau}\langle\mathcal{T}_\tau [\mathcal{O}_{q\nu}(\tau) \mathcal{O}_{q\nu^\prime}^\dagger (0)] \rangle
\end{align}
In the adiabatic limit $(i\omega_n \to 0)$, the self-energy determines the effective phonon Hamiltonian of Eq. (\ref{eq:Heffph}), where:
\begin{align}\label{eq:Nvv}
[\mathbb{N}_q]_{\nu,\nu^\prime} =  & \  \frac{1}{2} \lim_{i\omega_n \to 0}   \frac{i}{i\omega_n}\Pi_q^{\nu\nu^\prime}    (i\omega_n)
\end{align}
This can be expanded in terms of unperturbed electronic states $|n\rangle,|m\rangle$:
\begin{align}
\left[\mathbb{N}_q \right]_{\nu\nu^\prime} 
&  = 
\frac{i \hbar}{4}\sum_{nm} \frac{e^{-\beta E_n}-e^{-\beta E_m} }{Z_{\rm el}} \nonumber \\
& \left[  
        \frac{\langle n|  \mathcal{O}_{q\nu}^\dagger |m\rangle \langle m | \mathcal{O}_{q\nu^\prime}|n \rangle-\langle n|  \mathcal{O}_{q\nu^\prime} |m\rangle\langle m| \mathcal{O}_{q\nu}^\dagger |n \rangle}{(E_n-E_m)^2}\right]       
        \label{eqn:Iqv}
\end{align}
where $Z_{\rm el}$ is the unperturbed electronic partition function. This can be recognized as:
\begin{align}
\mathbb{N}_q = & \  \frac{i\hbar}{2}\langle \nabla_{\mathbf{u}_q^\dagger} \Psi_{\rm el}(\mathbf{u}) | \wedge | \nabla_{\mathbf{u}_q} \Psi_{\rm el}(\mathbf{u})\rangle = \frac{\hbar}{2}\Omega_q
\end{align}
where $\Omega_q$ is the ``nuclear Berry curvature'', which quantifies the geometric phase acquired by electronic wavefunctions under adiabatic motion of the atomic nuclei. As seen in Eq.~(\ref{eq:Heffph}), the nuclear Berry curvature couples to the phonon orbital momentum analogously to a magnetic field; its presence has various effects on vibrational dynamics \cite{mead1992geometric,
ren2024adiabatic}.  
 The full derivation of the above results is given in Supplementary Note 1. The relationship between $\mathbb{N}$ and the long-wavelength viscosity tensor is discussed in Supplementary Note 4.

It may be noted that various proposed {\it intrinsic} \cite{saito2019berry,ye2020phonon,ye2021phonon,zhang2021phonon} and {\it extrinsic} (impurity) \cite{sheng2006theory,guo2021extrinsic} mechanisms for the phonon THE can be understood within the Hall viscosity context. However, the adiabatic approximation does not include higher order skew-scattering vertex corrections or topological magnon-phonon crossings, which appear at finite energy. Nonetheless, we show that the adiabatic approximation is sufficient to capture the phenomenology of the phonon THE in \rucl.

\begin{figure*}[t]
\includegraphics[width=\linewidth]{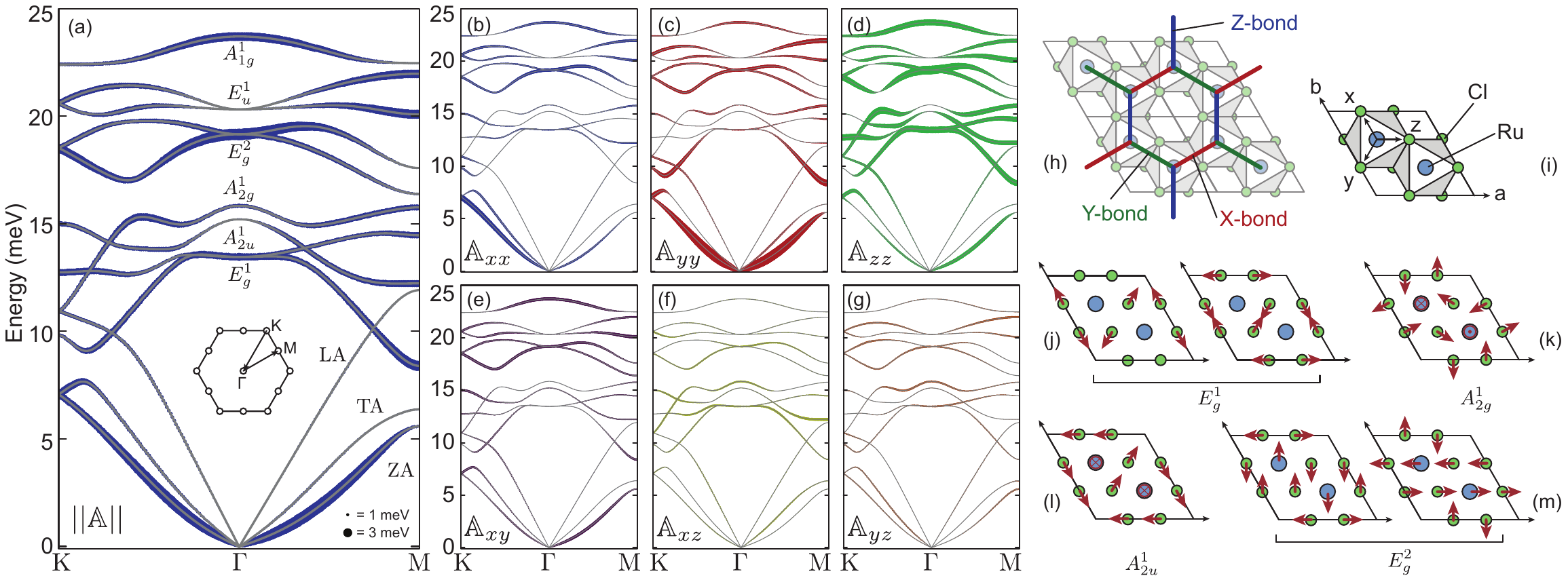}
\caption{{\bf Momentum dependence of the computed spin-phonon couplings.} $\mathbb{A}_{ij}^{q\nu}$ is depicted for the nearest-neighbor Z-bond along the high symmetry K-$\Gamma$-M path. (a): Frobenius norm $||\mathbb{A}||$. (b)-(g): Magnitude of different components of $\mathbb{A}$ in cubic coordinates. (h): Four unit cells of \rucl showing definition of nearest-neighbor X-, Y-, and Z-bonds. (i): Unit cell showing definition of cubic $x,y,z$ coordinates. (j)-(m): Primary displacements associated with $q=0$ optical phonons.}
\label{fig:bandresults}
\end{figure*}
\vspace{0.2cm}

{\bf RESULTS}

\vspace{0.2cm}

{\bf Spin-Phonon Couplings}

\vspace{0.2cm}

For magnetic insulators, two distinct spin-dependent {\it intrinsic} contributions to the Hall viscosity arise at lowest order. These are associated with spin-phonon couplings of the form \cite{dhakal2024mn}:
\begin{align}
    \mathcal{H}_{\rm sp-ph} = &  \sum_{q\nu} \bar{u}_{q\nu} \  \mathcal{A}_{q\nu}-\sum_{q\nu q^\prime \nu^\prime } \bar{u}_{q\nu} \bar{p}_{q^\prime \nu^\prime} \  \mathcal{L}_{q\nu ;q^\prime \nu^\prime} \label{eq:Hsp}
\end{align}
where $\bar{u}_{q\nu} = (a_{-q\nu}^\dagger + a_{q\nu})$ and $\bar{p}_{q\nu} = i(a_{-q\nu}^\dagger - a_{q\nu})$, and $a_{q\nu}^\dagger$ creates a phonon with momentum $q$ in band $\nu$. The phonon-operator units have been absorbed into $\mathcal{A}$ and $\mathcal{L}$, such that both operators have units of energy. For $\alpha$-RuCl$_3$, the $q$-space spin and bond operators are:
\begin{align}
\mathcal{A}_{q\nu} = & \ \frac{1}{\sqrt{N}}\sum_{ij} \left(\mathbf{S}_i \cdot \mathbb{A}_{ij}^{q\nu}\cdot \mathbf{S}_j\right) \ e^{-iq\cdot (r_i+r_j)/2}
\\ \label{eqn:Ldef}
\mathcal{L}_{q\nu ;q^\prime \nu^\prime} =& \  \frac{\hbar\omega_{q^\prime\nu^\prime}}{N}\sum_{i}\left(\mathbf{L}_{i}^{q\nu ;q^\prime \nu^\prime} \cdot \mathbf{S}_i\right)\ e^{-i(q+q^\prime)\cdot r_i}
\end{align}
where $N$ is the number of unit cells, and $\mathbf S_i$ describe the $j_{1/2}$ moments. The $\mathcal{A}$ operators parameterize the phonon-induced modulation of the intersite magnetic interactions, with $\mathbb{A}$ encoding the change of the $3\times 3$ magnetic coupling tensor. The $\mathcal{L}$ operators encode the two-phonon Raman coupling, with $\mathbf{L}$ a vector. This term arises from modulation of the local geometry around each Ru ion by the phonons, which alters the specific spin-orbital composition of the local moments \cite{orbach1961spin,capellmann1991spin}.

To assess the relative importance of these contributions, we estimate the spin-phonon couplings following the approach of \cite{dhakal2024mn}. Full details are given in the Supplementary Note 2. Briefly, we perform exact diagonalization of a coupled electron-phonon $d$-orbital Hamiltonian on each Ru site or bond of interest, and project the resulting low-energy Hamiltonians onto ideal $j_{1/2}$ states with different numbers of phonon quanta. The inclusion of phonons explicitly allows for the extraction of generic spin-phonon couplings in addition to the usual spin-spin couplings. We combine these results with {\it ab-initio} phonon calculations to obtain the full $q,\nu$-dependent spin-phonon Hamiltonian.

To reduce the complexity of the computations, we utilize a relaxed and symmetrized $P\bar{3}1m$ structure with AA stacking of adjacent layers, which facilitates the symmetry analysis below. In this structure, each Ru site has $D_3$ point group symmetry, and each nearest-neighbor bond has $C_{2h}$ symmetry.  
For the symmetrized structure, with reference to the spin Hamiltonian $\mathcal{H}_{\rm s} = \sum_{ij} \mathbf{S}_i \cdot \mathbb{J}_{ij} \cdot \mathbf{S}_j$, we compute the zeroth-order spin couplings as $(J,K,\Gamma,\Gamma^\prime) = (-2.79, -5.96, +3.06, -0.12)$\,meV, where:
\begin{align}\label{eqn:zbond}
\mathbb{J}_{\rm Z-bond}= \begin{pmatrix}
J & \Gamma & \Gamma^\prime \\
\Gamma & J & \Gamma^\prime \\
\Gamma^\prime & \Gamma^\prime & J+K
\end{pmatrix}
\end{align}
for the nearest-neighbor Z-bonds [Fig.~\ref{fig:bandresults}(h)]. The other bonds are related by symmetry \cite{rau2014generic,winter2016challenges}. 
The computed couplings are sufficiently close to reported bulk values \cite{
winter2017breakdown,eichstaedt2019deriving,kaib2022electronic,maksimov2020rethinking} to lend confidence to the {\it ab-initio} methodology.

In Fig.~\ref{fig:bandresults}(a-g), we first show the computed phonon dispersions and $\mathbb{A}_{ij}^{q\nu}$ couplings for the nearest-neighbor Z-bond. 
Despite consideration of the artificial $P\bar{3}1m$ structure, the phonon dispersion for in-plane momenta reproduces experimental measurements well \cite{
lebert2022acoustic}. 
Figure~\ref{fig:bandresults}(b)-(g) shows that the magnitudes of the different components of the spin-phonon bond operators are strongly momentum- and band-dependent. The coupling in each band can be rationalized by the corresponding real-space displacements shown in Fig.~\ref{fig:bandresults}(j)-(m). For example, the lowest optical mode at $q=0$, labelled $E_g^1$, primarily involves the symmetric motion of the Cl atoms bridging the nearest-neighbor bonds. For the Z-bond, this modulates the ligand-assisted hopping between $d_{yz}$ and $d_{xz}$ orbitals, which is the primary source of the Kitaev coupling $K$ \cite{rau2014generic,winter2016challenges}. As a consequence, $\mathbb{A}_{zz}$ is the largest component of the spin-phonon coupling on the Z-bond for that particular band, as shown in Fig.~\ref{fig:bandresults}(d). The optical $E_g^2$ mode involves a similar in-plane motion of the Cl atoms, but also a modulation of the Ru positions along the bond direction. The latter motion affects the direct $d_{xy}-d_{xy}$ hopping, which alters the $\Gamma$ and $J$ couplings. For this reason, finite contributions to $\mathbb{A}_{xx}$ and $\mathbb{A}_{xy}$ are apparent. 
While we leave full analysis of the optical-phonon couplings to future work, this discussion emphasizes that the competing exchange processes in $j_{1/2}$ systems generate {\it strongly anisotropic and band-dependent} spin-phonon couplings. 

For \rucl, the $\mathbf{L}$ couplings are found to be several orders of magnitude weaker than the $\mathbb{A}$ couplings, and are presented in full detail in Supplementary Note 3. Their weakness stems from the relatively strong spin-orbit coupling, which suppresses the mixing of the $j_{1/2}$ and $j_{3/2}$ states. 

We now discuss how each spin-phonon coupling contributes to the nuclear Berry curvature Eq.~\ref{eqn:Iqv}. The two-phonon Raman terms $\mathcal{L}$ arise from integrating out higher lying multiplet states, where $\mathcal{O}$ in Eq.~(\ref{eqn:Iqv}) are electron-phonon coupling operators, and $|m\rangle$ are excited $j_{3/2}$ spin-orbital multiplets. This coupling is featured prominently in phenomenological models of the phonon THE \cite{sheng2006theory,wang2009phonon,zhang2010topological}, where it is often approximated by $\mathcal{H}_{\rm eff} \sim \mathbf{S} \cdot (\mathbf{u} \times \mathbf{p})$. The $\mathcal{L}$-operators represent the generalization of these effects. They lead to a contribution to Hall viscosity of $[\mathbb{N}_q]_{\nu,\nu^\prime}\propto \frac{1}{N} \sum_i \langle \mathbf{L}_i^{-q\nu;q\nu^\prime} \cdot \mathbf{S}_i\rangle$, which is linear in a static $q=0$ magnetic order parameter such as the bulk magnetization (or altermagnetic order parameter if symmetry allows). This leads to the phonon equivalent of the anomalous Hall effect. However, it requires low-lying excited multiplets $|m\rangle$ for the phonons to couple to, which are not present in \rucl. We thus find that the ``Raman'' coupling mechanism is not relevant for the phonon THE in \rucl, which explains why experiments do not observe a contribution to $\kappa_{xy}$ that scales with the magnetization. 


\begin{figure*}[t]
\includegraphics[width=\linewidth]{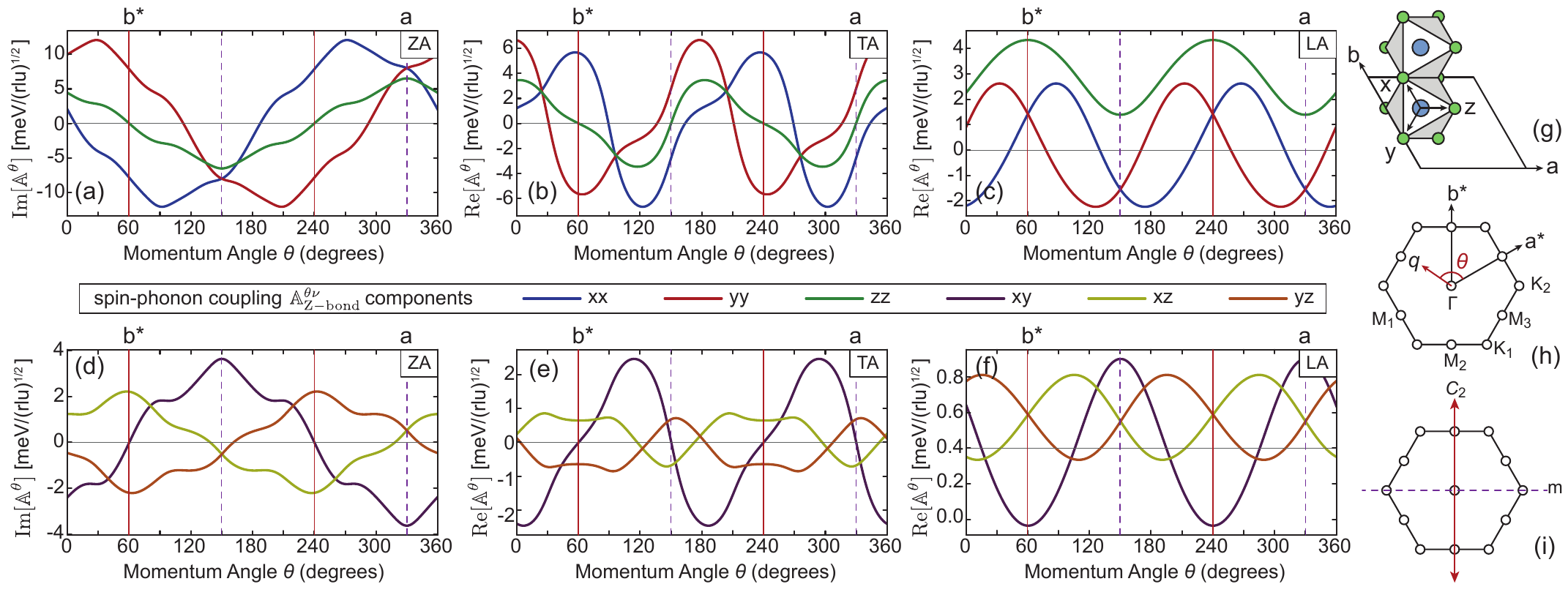}
\caption{{\bf Low-momentum angle dependence of acoustic-mode spin-phonon couplings.} 
Computed $\mathbb{A}_{ij}^{\theta \nu}$ for the nearest-neighbor Z-bond are shown as a function of the momentum angle $\theta$. (a)-(c): diagonal $\mathbb{A}_{xx}$, $\mathbb{A}_{yy}$, and $\mathbb{A}_{zz}$ components. (d)-(f): off-diagonal $\mathbb{A}_{xy}$, $\mathbb{A}_{xz}$, and $\mathbb{A}_{yz}$ components. The acoustic modes ZA, TA, and LA are labelled in the upper right of each panel. (g): Z-bond in real space showing definition of the unit cell $a,b$ axes and cubic ($x,y,z$) coordinates. (h): first Brillouin zone showing definition of the $a^*, b^*$ axes, and $\theta$ (measured from the $a^*$ axis). (i): orientation, in $q$-space, of the relevant symmetry operations for the Z-bond. 1 rlu = $4\pi/(\sqrt{3}a)$. 
}
\label{fig:Aq_results}
\end{figure*}


A second contribution to the nuclear Berry curvature arises from integrating out magnetic excitations within the lowest spin-orbital multiplet on each site. This produces a contribution to the Hall viscosity $\mathbb{N}$ where $\mathcal{O}$ in Eq.~(\ref{eqn:Iqv}) are the spin-phonon coupling operators $\mathcal{A}$ and $|m\rangle$ are excitations like magnons, spinons, or paramagnetic fluctuations of the $j_{1/2}$ moments. This leads to a Hall viscosity contribution related to antisymmetric dynamical bond-bond correlations $\langle\mathcal{T}_\tau [\mathcal{A}_{q\nu}(\tau) \mathcal{A}_{q\nu^\prime}^\dagger (0)] \rangle$ (see Supplementary Note 5). For \rucl, we find that this contribution dominates, as discussed below.

To address the low-temperature thermal transport properties, we focus on the acoustic phonon modes. These three modes nominally correspond to out-of-plane motion of the atoms transverse to the $q$-vector (labelled ZA), in-plane transverse motion (TA), and in-plane longitudinal motion (LA). In the long-wavelength limit $(q\to 0)$, assuming linear dispersion of the acoustic phonons, the spin-phonon couplings scale as:
\begin{align}
\lim_{q\to 0}\mathbb{A}_{ij}^{q\nu}\approx & \ \mathbb{A}_{ij}^{\theta\nu}\  |q|^{1/2} 
\label{eq:Aqnu_qapprox0}
\end{align}
where $\theta$ denotes the angle in $q$-space measured from the $a^*$-axis [Fig.~\ref{fig:Aq_results}(h)]. Note that there is gauge freedom in the complex phases of these couplings. As detailed in Supplementary Note 2, we choose a gauge by enforcing smoothness of the couplings at small finite $|q|$, which leads to $\text{Re}[\mathbb{A}_{ij}^{\theta;\rm ZA}] = \text{Im}[\mathbb{A}_{ij}^{\theta;\rm TA}] = \text{Im}[\mathbb{A}_{ij}^{\theta;\rm LA}] =0$.

Fig.~\ref{fig:Aq_results} shows the computed $\theta$-dependence of the $\mathbb{A}_{ij}^{\theta\nu}$ matrices for the nearest-neighbor Z-bond. At each $q$-point, time-reversal and inversion symmetries of the Hamiltonian ensure that $\mathbb{A}_{\alpha\beta} = \mathbb{A}_{\beta\alpha}$. In real space, the Z-bond is symmetric with respect to $C_2 \ || \ b^*$ and $m\perp b^*$ [Fig.~\ref{fig:Aq_results}(i)]. This restricts the spin-phonon couplings for certain high-symmetry momenta. For $q \ || \ b^*$ (parallel to the bond), the ZA and TA modes are odd with respect to the $C_2$, while the LA modes are even. For $q\perp b^*$ (perpendicular to the bond),  the TA modes are odd with respect to $m$, while the ZA and LA modes are even. 
Thus, symmetry requires some couplings to change sign as the momentum direction rotates around the Brillouin zone. For example [as seen in Fig.~\ref{fig:Aq_results}(a)], for the ZA mode: at $q \ ||\ a$, $\mathbb{A}_{xx}=\mathbb{A}_{yy}$, while at $q \ || \ b^*$, $\mathbb{A}_{xx}=-\mathbb{A}_{yy}$. The net result is that the spin-phonon couplings tend to wind around the origin of the Brillouin zone. Symmetry-enforced sign changes can be seen in most of the Z-bond spin-phonon couplings in Fig.~\ref{fig:Aq_results}(a-f).

The physical consequence of the strong $\theta$-dependence of $ \mathbb{A}_{ij}^{\theta\nu}$ can be understood as follows. In the effective phonon Hamiltonian Eq.~\ref{eq:Heffph}, the Hall viscosity $\mathbb{N}_q$ enters as an antisymmetric off-diagonal term that mixes phonon modes of differing polarization. As elaborated in \cite{ye2021phonon}, this leads to the acoustic Faraday effect, which has now been experimentally confirmed in $\alpha$-RuCl$_3$ \cite{shragai2025phonon}. Due to the $q$-dependence of $\mathbb{N}_q$, the resulting phonon eigenvectors can also wind around the Brillouin zone, which is reflected in a finite Berry curvature for the phonon bands, and thus a finite $\kappa_{xy}$. The finding that the $\mathcal{A}$ couplings have a $q$-dependence conducive to such winding suggests that finite phonon Berry curvature can arise from spin-phonon coupling under relatively generic conditions, as elaborated below. 

\vspace{0.2cm}

{\bf Phonon Longitudinal Thermal Conductivity}

\vspace{0.2cm}

\begin{figure}[t]
\includegraphics[width=\linewidth]{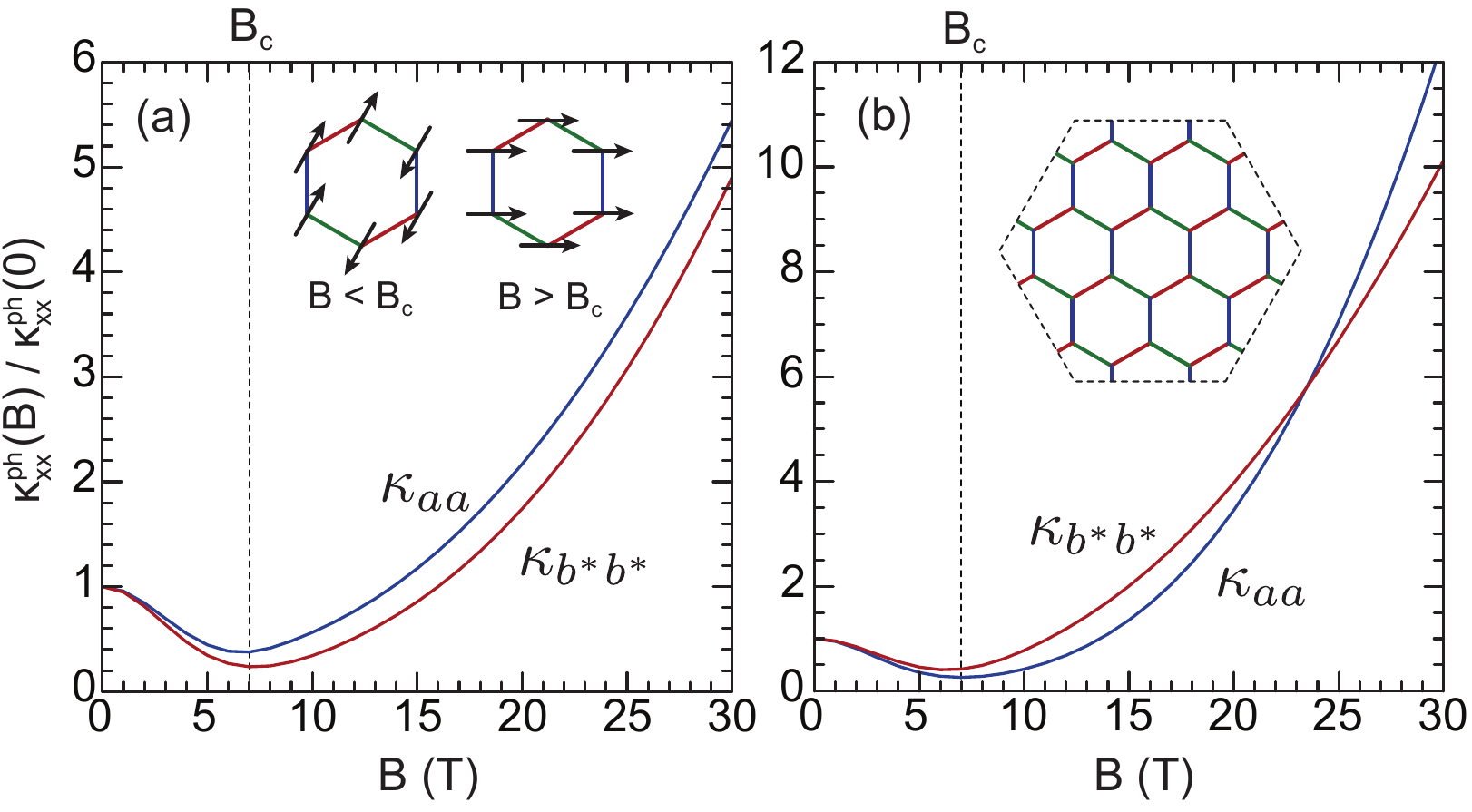}
\caption{ {\bf Longitudinal phonon thermal conductivity.} Shown for (a): magnetic field $\mathbf{B} \ || \ a$ and (b): $\mathbf{B} \ || \ b^*$. Inset in (a) shows zigzag ($B < B_c$) and polarized ($B > B_c$) magnetic states. Inset in (b) shows 24 site cluster used in ED.}
\label{fig:kxx_results}
\end{figure}

We now consider the effects of the computed spin-phonon couplings on the field-dependence of the low-temperature thermal transport. For this purpose, we employ exact diagonalization on the 24-site periodic cluster depicted in Fig.~\ref{fig:kxx_results}(b; inset) to evaluate the bond correlations appearing in the acoustic phonon self-energy [Eq.~(\ref{eq:phSE}) with $\mathcal{O}_{q\nu} = \mathcal{A}_{q\nu}$]. We consider the bare spin Hamiltonian with an in-plane magnetic field $\mathbf{B}$:
\begin{align}
\mathcal{H}_\mathrm{s} = \sum_{ij} \mathbf{S}_i \cdot \mathbb{J}_{ij} \cdot \mathbf{S}_j  - \mu_\mathrm{B} g_{ab}\sum_i \mathbf{B} \cdot \mathbf{S}_i
\end{align}
with the nearest-neighbor couplings computed for the relaxed structure, plus a third-neighbor Heisenberg coupling $J_3 = +0.8$\,meV, and an in-plane $g$-value of $g_{ab} = 2.3$. These values lie within the range reported in the literature \cite{maksimov2020rethinking}. 
For in-plane fields, this model has two phases [Fig.~\ref{fig:kxx_results}(a; inset)]. For $B<B_c\approx 7$\,T, it is nominally in an antiferromagnetic zigzag phase; for $B>B_c$ it adopts an asymptotically polarized phase. No intermediate phase is found.


Because ED does not resolve a fine $q$-dependence, we approximate $\mathcal{A}_{q\nu}|_{q\approx 0} \approx \frac{1}{\sqrt{N}} \sum_{ij} (\mathbf{S}_i \cdot \mathbb{A}_{ij}^{\theta\nu} \cdot \mathbf{S}_j)|q|^{1/2}$ [cf.~Eq.~\eqref{eq:Aqnu_qapprox0}]. This corresponds to retaining the $q$-dependence of the spin-phonon coupling operators, but approximating the relevant low-$q$ bond correlations by their $q\to 0$ limits. Within this approximation, the $q$-dependence of the phonon self-energy and Hall viscosity arises solely from the spin-phonon couplings.  The computed ED bond correlations strictly correspond to zero temperature but serve as a proxy for their low-temperature forms. 


\begin{figure*}[t]
\includegraphics[width=\linewidth]{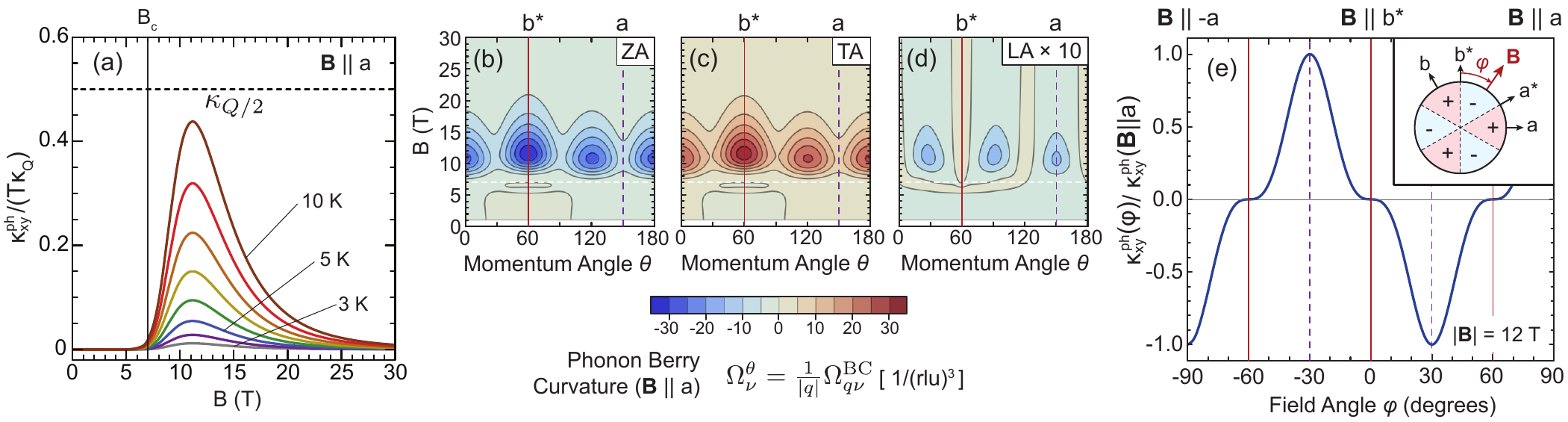}
\caption{ {\bf Phonon thermal Hall conductivity.} 
(a): Evolution of phonon thermal Hall conductivity $\kappa_{xy}^{\rm ph}$ as a function of field for $\mathbf{B}\ || \ a$ computed using ED bond correlations. 
(b-d): Phonon Berry curvature $\Omega_\nu^\theta$ as a function of field and momentum direction for three acoustic phonon bands; 1\,rlu = $4\pi/(\sqrt{3}a)$.
(e): Field angle dependence of $\kappa_{xy}^{\rm ph}$ at $|\mathbf{B}| = 12$\,T. Inset: Sign of $\kappa_{xy}^{\rm ph}$ as a function of in-plane field direction, with definition of field angle $\phi$. 
} \label{fig:kxy_results}
\end{figure*}


At fixed low temperature, assuming spin-phonon coupling is the dominant scattering mechanism, the in-plane longitudinal phonon thermal conductivity $\kappa_{\alpha\alpha}^{\rm ph}$ in the long-wavelength limit follows (Supplementary Note 5):
\begin{align}
    \kappa_{\alpha\alpha}^{\rm ph} \propto \sum_\nu \int_0^{2\pi}d\theta \ (\hat{q}\cdot\hat{\alpha})^2 \ \tau_\nu^\theta
\end{align}
where $(\hat{q}\cdot\hat{\alpha})$ denotes the projection of the momentum onto the transport direction, and $\tau_\nu^\theta$ is the band-dependent relative phonon lifetime, given by:
\begin{align}
\tau_\nu^\theta \equiv \frac{\hbar c_\nu |q|}{\text{Im}[\Pi_{q}^{\nu\nu}(\omega_{q\nu})]}
\end{align}
where $c_\nu = \partial \omega_{q\nu} / \partial|q|$ is the speed of sound for each band,
obtained from the {\it ab-initio} phonon calculations. Since $\text{Im}[\Pi_{q}^{\nu\nu}(\omega_{q\nu})] \propto |q|$, the relative lifetime depends only on the momentum direction. To compute $\kappa_{\alpha\alpha}$, the discrete poles of $\Pi_{q}^{\nu\nu}(\omega)$ were Lorentzian broadened with a width of $\gamma = 0.5$\,meV to extrapolate to low frequency. This introduces an unavoidable dependence of the computed phonon lifetimes on the broadening $\gamma$. For this reason, we plot $\kappa_{xx}^{\rm ph}$ relative to its zero-field value, which is insensitive to $\gamma$ (Supplemental Note 5).

Fig.~\ref{fig:kxx_results} shows the computed field-dependence of the in-plane $\kappa_{aa}^{\rm ph}$ and $\kappa_{b^* b^*}^{\rm ph}$ for $\mathbf{B} \ || \ a$ and $\mathbf{B} \ || \ b^*$. For both field directions and both propagation directions, the longitudinal thermal conductivity reaches a minimum near the critical field. This behavior is consistent with experimental measurements \cite{leahy2017anomalous,hentrich2018unusual,yu2018ultralow,kasahara2018majorana,yokoi2021half}, which report reductions of $\kappa_{\alpha\alpha}^{\rm ph}$ at $B_c$ by factors of 2--5 relative to the zero-field value. 
This behavior is readily understood. At both low field and high field, the spin excitations are gapped and well separated in energy from the low-$q$ acoustic phonons. At the critical field, the reduction of the spin gap implies an increased density of low-energy spin excitations, which enhances scattering. Since $\Pi_{q}^{\nu\nu}$ is a dynamical bond-bond correlation, the closure of a one-magnon gap at any $q$-point enhances scattering for phonons near $q=0$ due to contributions from multi-magnons at low-$q$. ED reproduces this effect, but does not capture a complete gap-closing due to finite size effects, resulting in a broader minimum in $\kappa_{\alpha\alpha}(B)$ than found in experiments. In the high-field phase, we find $\kappa_{\alpha\alpha}^{\rm ph}$ is larger for $\mathbf{B}\  ||\  b^*$ than for $\mathbf{B}\  ||\  a$ as a consequence of the details of the spin-phonon couplings. At all fields, $\kappa_{\alpha\alpha}^{\rm ph}$ is dominated by the LA phonons, which have the largest sound velocity and the weakest spin-phonon coupling (see Fig~\ref{fig:bandresults}).

\vspace{0.2cm}

{\bf Phonon Thermal Hall Conductivity}

\vspace{0.2cm}

To estimate the field-dependence of the thermal Hall conductivity $\kappa_{xy}^{\rm ph}$, we computed the Hall viscosity matrix $\mathbb{N}$ for the acoustic phonons in the zero-temperature limit by summing Eq.~(\ref{eq:Nvv}) over the ED poles without broadening. We then numerically evaluated the phonon Berry curvature $\Omega_{q\nu}^{\rm BC}$ by diagonalizing the phonon effective Hamiltonian [Eq.~(\ref{eq:Heffph})]. We find that the Berry curvature of the acoustic phonons is proportional to $|q|$ for small momenta, and thus define $\Omega_{\nu}^\theta = \Omega_{q\nu}^{\rm BC}/|q|$. Following \cite{ye2021phonon,qin2012berry,li2023magnons,katsura2010theory}:
\begin{align}\label{eq:kxyphmaintext}
    \frac{\kappa_{xy}^{\rm ph}}{T  \kappa_Q}\approx 
    - 2.00655   \sum_\nu \left( \frac{k_\mathrm{B}T}{\hbar c_\nu} \right)^3\int_0^{2\pi}  \Omega_\nu^\theta \ d\theta
\end{align}
where $\kappa_Q=\pi k_\mathrm{B}^2/(6\hbar)$ is the quantum of thermal conductance.

Fig.~\ref{fig:kxy_results}(a) shows the qualitative evolution of $\kappa_{xy}^{\rm ph}$ for $\mathbf{B} \ || \ a$.  Here, we have employed the $T = 0$ spin correlations to evaluate the phonon Berry curvature in Eq.~(\ref{eq:kxyphmaintext}), so that the depicted temperature dependence is approximate (as discussed further below). We find that $\kappa_{xy}^{\rm ph}$ rapidly increases above the critical field, reaches a maximum near 11\,T, and then decays with increasing field. This can be understood from the evolution of the phonon Berry curvature $\Omega_{\nu}^\theta$, depicted in Fig.~\ref{fig:kxy_results}(b-d). Above the critical field, there is a rapid increase in $\Omega_{\nu}^\theta$ for the ZA and TA phonons across all momentum directions (while the LA phonons develop nearly no Berry curvature). Ultimately,  $\kappa_{xy}^{\rm ph}$ is dominated by the ZA phonons, which have the smallest sound velocity (the largest thermal population), and develop the largest Berry curvature. An interesting consequence is that $\kappa_{xx}$ and $\kappa_{xy}$ do not probe the same phonons at low temperatures.

In Fig.~\ref{fig:kxy_results}(e), we show the computed dependence of $\kappa_{xy}^{\rm ph}$ on the direction of the in-plane field, for $|\mathbf{B}| = 12$ T. We find that the in-plane field-angle dependence agrees with experimental reports~\cite{czajka2023planar,yokoi2021half,imamura2024majorana}, with $\kappa_{xy}^{\rm ph}$ changing sign as the field is rotated through $a^*$, $b^*$, and symmetry-related directions.  While this mimics the expected angle-dependence of $\kappa_{xy}$ contributed by Majorana spinons in the pure Kitaev spin-liquid, it is also the simplest sign structure compatible with the symmetries of the crystal. 
For $\mathbf{B} \ || \ a^*, b^*$, symmetry requires $\kappa_{xy} = 0$ due to the twofold rotational symmetry  \cite{chern2021sign,kurumaji2023symmetry}. 
The overall sign of $\kappa_{xy}^{\rm ph}$ agrees with experiment, and is determined by the combined effect of the spin-phonon couplings and bond correlations. It is likely non-universal.

Together, these results reveal that a {\it planar} phonon THE is possible in \rucl due to {\it intrinsic} Hall viscosity. In contrast with the electronic Hall effect, an out-of-plane component of the magnetic field is not necessary, because the phonons couple to the emergent gauge field generated by the nuclear Berry curvature, not to the actual magnetic field. 
As we show in Supplementary Note 5, the Hall conductivity scales roughly as $\kappa_{xy}^{\rm ph} \propto \mathbb{N}^3 \sim (\mathcal{A}/\Delta_{s})^6$, where $\mathcal{A}$ is the spin-phonon coupling, and $\Delta_s$ is the spin-gap [see also Eq. (\ref{eq:Nvv}-\ref{eqn:Iqv})]. It is thus expected that the reduction of the spin-gap in the vicinity of $B_c$ enhances $\kappa_{xy}^{\rm ph}$ because the phonons are more strongly dressed by the low-lying spin excitations.
However, we find that the maximum in $\kappa_{xy}^{\rm ph}$ occurs above the critical field, in the asymptotically polarized phase (at 11\,T). This reflects the specific field dependence of the bond susceptibilities, which produces a peak in the Hall viscosities within the polarized phase. 
It may be noted that $\kappa_{xy}^{\rm ph}$ does not saturate at a constant value, as would be expected for dominant $\mathbf{L}$ couplings.

Lastly, we address the magnitude of $\kappa_{xy}^{\rm ph}$. Due to the sharp scaling noted above [$\kappa_{xy}^{\rm ph} \propto (\mathcal{A}/\Delta_{s})^6$], precise estimates of the magnitude require accurate modeling of the magnetic excitation spectrum and its temperature dependence. As noted above, we have estimated $\kappa_{xy}^{\rm ph}$ employing the zero-temperature spin correlations, which leads to $\kappa_{xy}^{\rm ph}/T \propto T^3$ due to the thermal population of phonons according to Eq.~(\ref{eq:kxyphmaintext}).
The predicted $\kappa_{xy}^{\rm ph}/T$ falls within an order of magnitude of $\frac{1}{2}\kappa_{Q}$, which agrees very well with experimental data. While a more precise calculation should account for the temperature dependence of the Hall viscosity in addition to the phonon population, we conclude that intrinsic phononic contributions are likely a significant portion of the observed $\kappa_{xy}$ in \rucl at low temperatures.

In summary, careful treatment of microscopic spin-phonon couplings reveals that the intrinsic phonon Hall effect is consistent with the low-temperature $\kappa_{xy}(B)$ in \rucl. This mechanism reproduces all essential experimental features including the field-dependence, sign, and order of magnitude. The intrinsic phonon THE considered here does not arise from real-space spin textures or non-trivial topology of the bare spin excitations. Instead, it arises from the momentum-dependence of the spin-phonon couplings, which can induce a finite phonon Berry curvature even if the electronic system is topologically trivial. A finite intrinsic Hall viscosity can likely be found in a variety of magnetic and non-magnetic insulators, for which differing phenomenology can be attributed to differences in the specific form of electron-phonon (or spin-phonon) coupling. The present results highlight a significant utility of first-principles spin-phonon coupling methods in disentangling the microscopic details underlying the phonon THE in specific materials.

\begin{acknowledgments}

\vspace{0.2cm}

{\bf ACKNOWLEDGEMENTS}

\vspace{0.2cm}

The authors acknowledge discussions with M. Ye, L. Mangeolle, L. Savary, L. Balents, A. Nevidomskyy, S. Ren, D. Vanderbilt, L. Taillefer, L. Chen, and Minhyea Lee. This research was funded by the Center for Functional Materials at WFU through a pilot grant, and Oak Ridge Associated Universities (ORAU) through the Ralph E. Powe Junior Faculty Enhancement Award to S.M.W. Part of the computations were performed using the Wake Forest University (WFU) High Performance Computing
Facility, a centrally managed computational resource available to WFU researchers including faculty, staff, students, and collaborators. This material is based upon work supported by the National Science Foundation under Grant No.~DMR-2338704.  D.A.S.K.\ and R.V.\ gratefully acknowledge support by the Deutsche Forschungsgemeinschaft (DFG, German Research Foundation) for funding through TRR 288—422213477 (projects A05 and B05). 
\end{acknowledgments}

\vspace{0.2cm}

{\bf AUTHOR CONTRIBUTIONS}

\vspace{0.2cm}

S.M.W. and R.V. conceived of the project. R.D., D.K., S.B. and S.M.W. performed the calculations. All authors contributed to the analysis and writing of the manuscript. S.M.W. and R.V. supervised the project.

\vspace{0.2cm}

{\bf COMPETING INTERESTS}

\vspace{0.2cm}

The authors declare no competing interests.

\bibliographystyle{naturemag}
\bibliography{rucl3spinphonon}

@article{shragai2025phonon,
  title={Phonon Hall Viscosity and the Intrinsic Thermal Hall Effect of {$\alpha$-RuCl$_3$}},
  author={Shragai, Avi and Horsley, Ezekiel and Kim, Subin and Kim, Young-June and Ramshaw, BJ},
  journal={arXiv preprint arXiv:2510.06443},
  year={2025}
}

@article{mook2014magnon,
  title={Magnon Hall effect and topology in kagome lattices: A theoretical investigation},
  author={Mook, Alexander and Henk, J{\"u}rgen and Mertig, Ingrid},
  journal={Phys. Rev. B},
  volume={89},
  number={13},
  pages={134409},
  year={2014},
  publisher={APS}
}

@article{nawwar2025large,
  title={Large thermal Hall effect in {MnPS$_3$}},
  author={Nawwar, Mohamed and Neumann, Robin R and Wen, Jiamin and Mertig, Ingrid and Mook, Alexander and Heremans, Joseph P},
  journal={Rep. Prog. Phys.},
  volume={88},
  number={8},
  pages={080503},
  year={2025},
  publisher={IOP Publishing}
}

@article{mead1992geometric,
  title={The geometric phase in molecular systems},
  author={Mead, C Alden},
  journal={Rev. Mod. Phys.},
  volume={64},
  number={1},
  pages={51},
  year={1992},
  publisher={APS}
}

@article{birss1962property,
  title={Property tensors in magnetic crystal classes},
  author={Birss, RR},
  journal={Proceedings of the Physical Society},
  volume={79},
  number={5},
  pages={946},
  year={1962},
  publisher={IOP Publishing}
}

@article{valenti1999novel,
  title={Novel nonreciprocal acoustic effects in antiferromagnets},
  author={Valenti, Roser and Gros, Claudius and Muthukumar, VN},
  journal={Europhysics Letters},
  volume={45},
  number={2},
  pages={242},
  year={1999},
  publisher={IOP Publishing}
}

@article{dhakal2024mn,
  title={Spin-Phonon Coupling in Transition Metal Insulators: General Computational Approach and Application to {MnPSe$_3$}},
  author={Dhakal, Ramesh and Griffith, Samuel and Choi, Kate and Winter, Stephen M},
  journal={arXiv preprint arXiv:2407.00659},
  year={2024}
}

@article{singh2024phonon,
  title={Phonon dynamics in the chiral Kitaev spin liquid},
  author={Singh, Susmita and Stavropoulos, P Peter and Perkins, Natalia B},
  journal={Phys. Rev. B},
  volume={110},
  number={9},
  pages={094431},
  year={2024},
  publisher={APS}
}

@article{kaib2022electronic,
  title={Electronic and magnetic properties of the {RuX$_3$} {(X = Cl, Br, I)} family: two siblings—and a cousin?},
  author={Kaib, David AS and Riedl, Kira and Razpopov, Aleksandar and Li, Ying and Backes, Steffen and Mazin, Igor I and Valent{\'\i}, Roser},
  journal={npj Quantum Mater.},
  volume={7},
  number={1},
  pages={75},
  year={2022},
  publisher={Nature Publishing Group UK London}
}

@article{kurumaji2023symmetry,
  title={Symmetry-based requirement for the measurement of electrical and thermal Hall conductivity under an in-plane magnetic field},
  author={Kurumaji, Takashi},
  journal={Phys. Rev. Research},
  volume={5},
  number={2},
  pages={023138},
  year={2023},
  publisher={APS}
}

@article{kitaev2006anyons,
  title={Anyons in an exactly solved model and beyond},
  author={Kitaev, Alexei},
  journal={Annals of Physics},
  volume={321},
  number={1},
  pages={2--111},
  year={2006},
  publisher={Elsevier}
}

@article{kane1997quantized,
  title={Quantized thermal transport in the fractional quantum Hall effect},
  author={Kane, CL and Fisher, Matthew PA},
  journal={Phys. Rev. B},
  volume={55},
  number={23},
  pages={15832},
  year={1997},
  publisher={APS}
}

@article{banerjee2018observation,
  title={Observation of half-integer thermal Hall conductance},
  author={Banerjee, Mitali and Heiblum, Moty and Umansky, Vladimir and Feldman, Dima E and Oreg, Yuval and Stern, Ady},
  journal={Nature},
  volume={559},
  number={7713},
  pages={205--210},
  year={2018},
  publisher={Nature Publishing Group UK London}
}

@article{saito2019berry,
  title={Berry phase of phonons and thermal Hall effect in nonmagnetic insulators},
  author={Saito, Takuma and Misaki, Kou and Ishizuka, Hiroaki and Nagaosa, Naoto},
  journal={Phys. Rev. Lett.},
  volume={123},
  number={25},
  pages={255901},
  year={2019},
  publisher={APS}
}

@article{flebus2022charged,
  title={Charged defects and phonon Hall effects in ionic crystals},
  author={Flebus, Benedetta and MacDonald, AH},
  journal={Phys. Rev. B},
  volume={105},
  number={22},
  pages={L220301},
  year={2022},
  publisher={APS}
}

@article{flebus2023phonon,
  title={Phonon Hall viscosity of ionic crystals},
  author={Flebus, B and MacDonald, AH},
  journal={Phys. Rev. Lett.},
  volume={131},
  number={23},
  pages={236301},
  year={2023},
  publisher={APS}
}

@article{schonemann2020ThermalMagnetoelasticPropertiesb,
  title = {Thermal and Magnetoelastic Properties of {$\alpha$-RuCl$_3$} in the Field-Induced Low-Temperature States},
  author = {Sch{\"o}nemann, Rico and Imajo, Shusaku and Weickert, Franziska and Yan, Jiaqiang and Mandrus, David G. and Takano, Yasumasa and Brosha, Eric L. and Rosa, Priscila F. S. and Nagler, Stephen E. and Kindo, Koichi and Jaime, Marcelo},
  year = {2020},
  month = dec,
  journal = {Phys. Rev. B},
  volume = {102},
  number = {21},
  pages = {214432},
  issn = {2469-9950, 2469-9969},
  doi = {10.1103/PhysRevB.102.214432},
  urldate = {2021-03-04},
  langid = {english}
}

@article{gass2020FieldinducedTransitionsKitaev,
  title = {Field-Induced Transitions in the Kitaev Material {$\alpha$-{R}u{C}l$_3$} Probed by Thermal Expansion and Magnetostriction},
  author = {Gass, S. and C{\^o}nsoli, P. M. and Kocsis, V. and Corredor, L. T. and {Lampen-Kelley}, P. and Mandrus, D. G. and Nagler, S. E. and Janssen, L. and Vojta, M. and B{\"u}chner, B. and Wolter, A. U. B.},
  year = {2020},
  month = jun,
  journal = {Phys. Rev. B},
  volume = {101},
  number = {24},
  pages = {245158},
  publisher = {American Physical Society},
  doi = {10.1103/PhysRevB.101.245158},
  urldate = {2020-06-30}
}

@article{kocsis2022MagnetoelasticCouplingAnisotropy,
  title = {Magnetoelastic Coupling Anisotropy in the {{Kitaev}} Material {$\alpha$-RuCl$_3$}},
  author = {Kocsis, Vilmos and Kaib, David A. S. and Riedl, Kira and Gass, Sebastian and {Lampen-Kelley}, Paula and Mandrus, David G. and Nagler, Stephen E. and P{\'e}rez, Nicol{\'a}s and Nielsch, Kornelius and B{\"u}chner, Bernd and Wolter, Anja U. B. and Valent{\'i}, Roser},
  year = {2022},
  month = mar,
  journal = {Phys. Rev. B},
  volume = {105},
  number = {9},
  pages = {094410},
  issn = {2469-9950, 2469-9969},
  doi = {10.1103/PhysRevB.105.094410},
  urldate = {2022-03-07},
  langid = {english}
}

@article{winter2016challenges,
  title={Challenges in design of Kitaev materials: Magnetic interactions from competing energy scales},
  author={Winter, Stephen M and Li, Ying and Jeschke, Harald O and Valent{\'\i}, Roser},
  journal={Phys. Rev. B},
  volume={93},
  number={21},
  pages={214431},
  year={2016},
  publisher={APS}
}

@article{winter2017breakdown,
  title={Breakdown of magnons in a strongly spin-orbital coupled magnet},
  author={Winter, Stephen M and Riedl, Kira and Maksimov, Pavel A and Chernyshev, Alexander L and Honecker, Andreas and Valent{\'\i}, Roser},
  journal={	Nat. Commun.},
  volume={8},
  number={1},
  pages={1152},
  year={2017},
  publisher={Nature Publishing Group UK London}
}

@article{maksimov2020rethinking,
  title={Rethinking {$\alpha$-RuCl$_3$}},
  author={Maksimov, PA and Chernyshev, AL},
  journal={Phys. Rev. Research},
  volume={2},
  number={3},
  pages={033011},
  year={2020},
  publisher={APS}
}

@article{orbach1961spin,
  title={Spin-lattice relaxation in rare-earth salts},
  author={Orbach, R},
  journal={Proc. R. soc. Lond. Ser. A-Contain. Pap. Math. Phys.},
  volume={264},
  number={1319},
  pages={458--484},
  year={1961},
  publisher={The Royal Society London}
}

@article{capellmann1991spin,
  title={Spin-phonon coupling in intermediate valency: exactly solvable models},
  author={Capellmann, H and Lipinski, S},
  journal={Z. Phys. B: Condens. Matter},
  volume={83},
  number={2},
  pages={199--205},
  year={1991},
  publisher={Springer}
}

@article{ren2024adiabatic,
  title={Adiabatic dynamics of coupled spins and phonons in magnetic insulators},
  author={Ren, Shang and Bonini, John and Stengel, Massimiliano and Dreyer, Cyrus E and Vanderbilt, David},
  journal={Phys. Rev. X},
  volume={14},
  number={1},
  pages={011041},
  year={2024},
  publisher={APS}
}

@article{zhang2024thermal,
  title={Thermal Hall effects in quantum magnets},
  author={Zhang, Xiao-Tian and Gao, Yong Hao and Chen, Gang},
  journal={Phys. Rep.},
  volume={1070},
  pages={1--59},
  year={2024},
  publisher={Elsevier}
}

@article{mcclarty2018topological,
  title={Topological magnons in Kitaev magnets at high fields},
  author={McClarty, PA and Dong, X-Y and Gohlke, M and Rau, JG and Pollmann, F and Moessner, R and Penc, K},
  journal={Phys. Rev. B},
  volume={98},
  number={6},
  pages={060404},
  year={2018},
  publisher={APS}
}

@article{zhang2021topological,
  title={Topological magnons for thermal Hall transport in frustrated magnets with bond-dependent interactions},
  author={Zhang, Emily Z and Chern, Li Ern and Kim, Yong Baek},
  journal={Phys. Rev. B},
  volume={103},
  number={17},
  pages={174402},
  year={2021},
  publisher={APS}
}

@article{chern2021sign,
  title={Sign structure of thermal Hall conductivity and topological magnons for in-plane field polarized Kitaev magnets},
  author={Chern, Li Ern and Zhang, Emily Z and Kim, Yong Baek},
  journal={Phys. Rev. Lett.},
  volume={126},
  number={14},
  pages={147201},
  year={2021},
  publisher={APS}
}

@article{ye2020phonon,
  title={Phonon dynamics in the Kitaev spin liquid},
  author={Ye, Mengxing and Fernandes, Rafael M and Perkins, Natalia B},
  journal={Phys. Rev. Research},
  volume={2},
  number={3},
  pages={033180},
  year={2020},
  publisher={APS}
}

@article{ye2021phonon,
  title={Phonon Hall viscosity in magnetic insulators},
  author={Ye, Mengxing and Savary, Lucile and Balents, Leon},
  journal={arXiv preprint arXiv:2103.04223},
  year={2021}
}

@article{barkeshli2012dissipationless,
  title={Dissipationless phonon Hall viscosity},
  author={Barkeshli, Maissam and Chung, Suk Bum and Qi, Xiao-Liang},
  journal={Phys. Rev. B},
  volume={85},
  number={24},
  pages={245107},
  year={2012},
  publisher={APS}
}

@article{mangeolle2022phonon,
  title={Phonon thermal hall conductivity from scattering with collective fluctuations},
  author={Mangeolle, L{\'e}o and Balents, Leon and Savary, Lucile},
  journal={Phys. Rev. X},
  volume={12},
  number={4},
  pages={041031},
  year={2022},
  publisher={APS}
}

@article{mangeolle2022thermal,
  title={Thermal conductivity and theory of inelastic scattering of phonons by collective fluctuations},
  author={Mangeolle, L{\'e}o and Balents, Leon and Savary, Lucile},
  journal={Phys. Rev. B},
  volume={106},
  number={24},
  pages={245139},
  year={2022},
  publisher={APS}
}

@article{mori2014origin,
  title={Origin of the phonon Hall effect in rare-earth garnets},
  author={Mori, Michiyasu and Spencer-Smith, Alexander and Sushkov, Oleg P and Maekawa, Sadamichi},
  journal={Phys. Rev. Lett.},
  volume={113},
  number={26},
  pages={265901},
  year={2014},
  publisher={APS}
}

@article{sun2022large,
  title={Large extrinsic phonon thermal Hall effect from resonant scattering},
  author={Sun, Xiao-Qi and Chen, Jing-Yuan and Kivelson, Steven A},
  journal={Phys. Rev. B},
  volume={106},
  number={14},
  pages={144111},
  year={2022},
  publisher={APS}
}

@article{guo2021extrinsic,
  title={Extrinsic phonon thermal Hall transport from Hall viscosity},
  author={Guo, Haoyu and Sachdev, Subir},
  journal={Phys. Rev. B},
  volume={103},
  number={20},
  pages={205115},
  year={2021},
  publisher={APS}
}

@article{guo2022resonant,
  title={Resonant thermal Hall effect of phonons coupled to dynamical defects},
  author={Guo, Haoyu and Joshi, Darshan G and Sachdev, Subir},
  journal={Proc. Natl. Acad. Sci. U.S.A.},
  volume={119},
  number={46},
  pages={e2215141119},
  year={2022},
  publisher={National Acad Sciences}
}

@article{sheng2006theory,
  title={Theory of the phonon Hall effect in paramagnetic dielectrics},
  author={Sheng, L and Sheng, DN and Ting, CS},
  journal={Phys. Rev. Lett.},
  volume={96},
  number={15},
  pages={155901},
  year={2006},
  publisher={APS}
}

@article{wang2009phonon,
  title={Phonon Hall thermal conductivity from the Green-Kubo formula},
  author={Wang, Jian-Sheng and Zhang, Lifa},
  journal={Phys. Rev. B},
  volume={80},
  number={1},
  pages={012301},
  year={2009},
  publisher={APS}
}

@article{zhang2010topological,
  title={Topological nature of the phonon Hall effect},
  author={Zhang, Lifa and Ren, Jie and Wang, Jian-Sheng and Li, Baowen},
  journal={Phys. Rev. Lett.},
  volume={105},
  number={22},
  pages={225901},
  year={2010},
  publisher={APS}
}

@article{zhang2021phonon,
  title={Phonon Hall viscosity from phonon-spinon interactions},
  author={Zhang, Yunchao and Teng, Yanting and Samajdar, Rhine and Sachdev, Subir and Scheurer, Mathias S},
  journal={Phys. Rev. B},
  volume={104},
  number={3},
  pages={035103},
  year={2021},
  publisher={APS}
}

@article{zhang2019thermal,
  title={Thermal hall effect induced by magnon-phonon interactions},
  author={Zhang, Xiaoou and Zhang, Yinhan and Okamoto, Satoshi and Xiao, Di},
  journal={Phys. Rev. Lett.},
  volume={123},
  number={16},
  pages={167202},
  year={2019},
  publisher={APS}
}

@article{nasu2017thermal,
  title={Thermal transport in the Kitaev model},
  author={Nasu, Joji and Yoshitake, Junki and Motome, Yukitoshi},
  journal={Phys. Rev. Lett.},
  volume={119},
  number={12},
  pages={127204},
  year={2017},
  publisher={APS}
}

@article{katsura2010theory,
  title={Theory of the thermal Hall effect in quantum magnets},
  author={Katsura, Hosho and Nagaosa, Naoto and Lee, Patrick A},
  journal={Phys. Rev. Lett.},
  volume={104},
  number={6},
  pages={066403},
  year={2010},
  publisher={APS}
}

@article{qin2012berry,
  title={Berry curvature and the phonon Hall effect},
  author={Qin, Tao and Zhou, Jianhui and Shi, Junren},
  journal={Phys. Rev. B},
  volume={86},
  number={10},
  pages={104305},
  year={2012},
  publisher={APS}
}

@article{thingstad2019chiral,
  title={Chiral phonon transport induced by topological magnons},
  author={Thingstad, Even and Kamra, Akashdeep and Brataas, Arne and Sudb{\o}, Asle},
  journal={Phys. Rev. Lett.},
  volume={122},
  number={10},
  pages={107201},
  year={2019},
  publisher={APS}
}

@article{huang2021topological,
  title={Topological phonon-magnon hybrid excitations in a two-dimensional honeycomb ferromagnet},
  author={Huang, Han and Tian, Zhiting},
  journal={Phys. Rev. B},
  volume={104},
  number={6},
  pages={064305},
  year={2021},
  publisher={APS}
}

@article{toth2015linear,
  title={Linear spin wave theory for single-Q incommensurate magnetic structures},
  author={Toth, S and Lake, B},
  journal={	J. Phys. Condens. Matter},
  volume={27},
  number={16},
  pages={166002},
  year={2015},
  publisher={IOP Publishing}
}

@article{matsumoto2014thermal,
  title={Thermal Hall effect of magnons in magnets with dipolar interaction},
  author={Matsumoto, Ryo and Shindou, Ryuichi and Murakami, Shuichi},
  journal={Phys. Rev. B},
  volume={89},
  number={5},
  pages={054420},
  year={2014},
  publisher={APS}
}

@article{kondo2020non,
  title={Non-Hermiticity and topological invariants of magnon Bogoliubov--de Gennes systems},
  author={Kondo, Hiroki and Akagi, Yutaka and Katsura, Hosho},
  journal={Progress of Theoretical and Experimental Physics},
  volume={2020},
  number={12},
  pages={12A104},
  year={2020},
  publisher={Oxford University Press}
}

@article{rau2014generic,
  title={Generic spin model for the honeycomb iridates beyond the Kitaev limit},
  author={Rau, Jeffrey G and Lee, Eric Kin-Ho and Kee, Hae-Young},
  journal={Phys. Rev. Lett.},
  volume={112},
  number={7},
  pages={077204},
  year={2014},
  publisher={APS}
}

@article{li2023magnons,
  title={Magnons, phonons, and thermal Hall effect in the candidate Kitaev magnet {$\alpha$-RuCl$_3$}},
  author={Li, Shuyi and Yan, Han and Nevidomskyy, Andriy H},
  journal={Phys. Rev. B},
  volume={112},
  number={13},
  pages={134413},
  year={2025},
  publisher={APS}
}

@article{li2022thermal,
  title={Thermal Hall effect in the Kitaev-Heisenberg system with spin-phonon coupling},
  author={Li, Shaozhi and Okamoto, Satoshi},
  journal={Phys. Rev. B},
  volume={106},
  number={2},
  pages={024413},
  year={2022},
  publisher={APS}
}

@article{hentrich2018unusual,
  title={Unusual phonon heat transport in {$\alpha$-RuCl$_3$}: strong spin-phonon scattering and field-induced spin gap},
  author={Hentrich, Richard and Wolter, Anja UB and Zotos, Xenophon and Brenig, Wolfram and Nowak, Domenic and Isaeva, Anna and Doert, Thomas and Banerjee, Arnab and Lampen-Kelley, Paula and Mandrus, David G and others},
  journal={Phys. Rev. Lett.},
  volume={120},
  number={11},
  pages={117204},
  year={2018},
  publisher={APS}
}

@article{leahy2017anomalous,
  title={Anomalous thermal conductivity and magnetic torque response in the honeycomb magnet {$\alpha$-RuCl$_3$}},
  author={Leahy, Ian A and Pocs, Christopher A and Siegfried, Peter E and Graf, David and Do, S-H and Choi, Kwang-Yong and Normand, B and Lee, Minhyea},
  journal={Phys. Rev. Lett.},
  volume={118},
  number={18},
  pages={187203},
  year={2017},
  publisher={APS}
}

@article{yu2018ultralow,
  title={Ultralow-Temperature Thermal Conductivity of the Kitaev Honeycomb Magnet {$\alpha$-RuCl$_3$} across the Field-Induced Phase Transition},
  author={Yu, YJ and Xu, Yang and Ran, KJ and Ni, JM and Huang, YY and Wang, JH and Wen, JS and Li, SY},
  journal={Phys. Rev. Lett.},
  volume={120},
  number={6},
  pages={067202},
  year={2018},
  publisher={APS}
}

@article{kasahara2018unusual,
  title={Unusual thermal Hall effect in a Kitaev spin liquid candidate {$\alpha$-RuCl$_3$}},
  author={Kasahara, Y and Sugii, K and Ohnishi, T and Shimozawa, M and Yamashita, M and Kurita, N and Tanaka, H and Nasu, J and Motome, Y and Shibauchi, T and Matsuda, Y},
  journal={Phys. Rev. Lett.},
  volume={120},
  number={21},
  pages={217205},
  year={2018},
  publisher={APS}
}

@article{kasahara2018majorana,
  title={Majorana quantization and half-integer thermal quantum Hall effect in a Kitaev spin liquid},
  author={Kasahara, Y and Ohnishi, T and Mizukami, Y and Tanaka, O and Ma, Sixiao and Sugii, K and Kurita, N and Tanaka, H and Nasu, J and Motome, Y and others},
  journal={Nature},
  volume={559},
  number={7713},
  pages={227--231},
  year={2018},
  publisher={Nature Publishing Group UK London}
}

@article{imamura2024majorana,
  title={Majorana-fermion origin of the planar thermal Hall effect in the Kitaev magnet {$\alpha$-RuCl$_3$}},
  author={Imamura, Kumpei and Suetsugu, Shota and Mizukami, Yuta and Yoshida, Yusei and Hashimoto, Kenichiro and Ohtsuka, Kenichi and Kasahara, Yuichi and Kurita, Nobuyuki and Tanaka, Hidekazu and Noh, Pureum and others},
  journal={Sci. Adv.},
  volume={10},
  number={11},
  pages={eadk3539},
  year={2024},
  publisher={American Association for the Advancement of Science}
}

@article{hentrich2019large,
  title={Large thermal Hall effect in {$\alpha$-RuCl$_3$}: Evidence for heat transport by Kitaev-Heisenberg paramagnons},
  author={Hentrich, Richard and Roslova, Maria and Isaeva, Anna and Doert, Thomas and Brenig, Wolfram and B{\"u}chner, Bernd and Hess, Christian},
  journal={Phys. Rev. B},
  volume={99},
  number={8},
  pages={085136},
  year={2019},
  publisher={APS}
}

@article{yokoi2021half,
  title={Half-integer quantized anomalous thermal Hall effect in the Kitaev material candidate {$\alpha$-RuCl$_3$}},
  author={Yokoi, T and Ma, S and Kasahara, Y and Kasahara, S and Shibauchi, T and Kurita, N and Tanaka, H and Nasu, J and Motome, Y and Hickey, C and others},
  journal={Science},
  volume={373},
  number={6554},
  pages={568--572},
  year={2021},
  publisher={American Association for the Advancement of Science}
}

@article{lefranccois2022evidence,
  title={Evidence of a Phonon Hall Effect in the Kitaev Spin Liquid Candidate {$\alpha$-RuCl$_3$}},
  author={Lefran{\c{c}}ois, {\'E} and Grissonnanche, G and Baglo, J and Lampen-Kelley, P and Yan, J-Q and Balz, C and Mandrus, D and Nagler, SE and Kim, S and Kim, Young-June and others},
  journal={Phys. Rev. X},
  volume={12},
  number={2},
  pages={021025},
  year={2022},
  publisher={APS}
}

@article{bruin2022robustness,
  title={Robustness of the thermal Hall effect close to half-quantization in {$\alpha$-RuCl$_3$}},
  author={Bruin, JAN and Claus, RR and Matsumoto, Y and Kurita, N and Tanaka, H and Takagi, H},
  journal={	Nat. Phys.},
  volume={18},
  number={4},
  pages={401--405},
  year={2022},
  publisher={Nature Publishing Group UK London}
}

@article{czajka2023planar,
  title={Planar thermal Hall effect of topological bosons in the Kitaev magnet {$\alpha$-RuCl$_3$}},
  author={Czajka, Peter and Gao, Tong and Hirschberger, Max and Lampen-Kelley, Paula and Banerjee, Arnab and Quirk, Nicholas and Mandrus, David G and Nagler, Stephen E and Ong, N Phuan},
  journal={Nat. Mater.},
  volume={22},
  number={1},
  pages={36--41},
  year={2023},
  publisher={Nature Publishing Group UK London}
}

@article{zhang2024stacking,
  title={Stacking disorder and thermal transport properties of {$\alpha$-RuCl$_3$}},
  author={Zhang, Heda and McGuire, Michael A and May, Andrew F and Chao, Hsin-Yun and Zheng, Qiang and Chi, Miaofang and Sales, Brian C and Mandrus, David G and Nagler, Stephen E and Miao, Hu and others},
  journal={Phys. Rev. Mater.},
  volume={8},
  number={1},
  pages={014402},
  year={2024},
  publisher={APS}
}

@article{sandilands2016spin,
  title={Spin-orbit excitations and electronic structure of the putative Kitaev magnet {$\alpha$-RuCl$_3$}},
  author={Sandilands, Luke J and Tian, Yao and Reijnders, Anjan A and Kim, Heung-Sik and Plumb, Kemp W and Kim, Young-June and Kee, Hae-Young and Burch, Kenneth S},
  journal={Phys. Rev. B},
  volume={93},
  number={7},
  pages={075144},
  year={2016},
  publisher={APS}
}

@article{sandilands2016optical,
  title={Optical probe of Heisenberg-Kitaev magnetism in {$\alpha$-RuCl$_3$}},
  author={Sandilands, Luke J and Sohn, Chang Hee and Park, Hyun Ju and Kim, So Yeun and Kim, Kyung Wan and Sears, Jennifer A and Kim, Young-June and Noh, Tae Won},
  journal={Phys. Rev. B},
  volume={94},
  number={19},
  pages={195156},
  year={2016},
  publisher={APS}
}

@article{lebert2023nonlocal,
  title={Nonlocal features of the spin-orbit exciton in Kitaev materials},
  author={Lebert, Blair W and Kim, Subin and Kim, Beom Hyun and Chun, Sae Hwan and Casa, Diego and Choi, Jaewon and Agrestini, Stefano and Zhou, Kejin and Garcia-Fernandez, Mirian and Kim, Young-June},
  journal={Phys. Rev. B},
  volume={108},
  number={15},
  pages={155122},
  year={2023},
  publisher={APS}
}

@book{sugano1970multiplets,
editor = {Satoru Sugano and Yukito Tanabe and Hiroshi Kamimura},
series = {Pure and Applied Physics},
publisher = {Elsevier},
volume = {33},
pages = {294-301},
year = {1970},
title = {Multiplets of Transition-Metal Ions in Crystals},
issn = {0079-8193},
doi = {https://doi.org/10.1016/B978-0-12-676050-7.50020-2}
}

@article{togo2023first,
  title={First-principles phonon calculations with phonopy and phono3py},
  author={Togo, Atsushi},
  journal={	J. Phys. Soc. Jpn.},
  volume={92},
  number={1},
  pages={012001},
  year={2023},
  publisher={The Physical Society of Japan}
}

@article{opahle1999full,
  title={Full-potential band-structure calculation of iron pyrite},
  author={Opahle, I and Koepernik, K and Eschrig, H},
  journal={Phys. Rev. B},
  volume={60},
  number={20},
  pages={14035},
  year={1999},
  publisher={APS}
}

@article{perdew1996generalized,
  title={Generalized gradient approximation made simple},
  author={Perdew, John P and Burke, Kieron and Ernzerhof, Matthias},
  journal={Phys. Rev. Lett.},
  volume={77},
  number={18},
  pages={3865},
  year={1996},
  publisher={APS}
}

@article{koepernik2023symmetry,
  title={Symmetry-conserving maximally projected Wannier functions},
  author={Koepernik, K and Janson, O and Sun, Yan and Van Den Brink, J},
  journal={Phys. Rev. B},
  volume={107},
  number={23},
  pages={235135},
  year={2023},
  publisher={APS}
}

@article{eichstaedt2019deriving,
  title={Deriving models for the Kitaev spin-liquid candidate material {$\alpha$-RuCl$_3$} from first principles},
  author={Eichstaedt, Casey and Zhang, Yi and Laurell, Pontus and Okamoto, Satoshi and Eguiluz, Adolfo G and Berlijn, Tom},
  journal={Phys. Rev. B},
  volume={100},
  number={7},
  pages={075110},
  year={2019},
  publisher={APS}
}

@article{kresse1996efficient,
  title={Efficient iterative schemes for ab initio total-energy calculations using a plane-wave basis set},
  author={Kresse, Georg and Furthm{\"u}ller, J{\"u}rgen},
  journal={Phys. Rev. B},
  volume={54},
  number={16},
  pages={11169},
  year={1996},
  publisher={APS}
}

@article{kresse1996efficiency,
  title={Efficiency of ab-initio total energy calculations for metals and semiconductors using a plane-wave basis set},
  author={Kresse, Georg and Furthm{\"u}ller, J{\"u}rgen},
  journal={Comput. Mater. Sci.},
  volume={6},
  number={1},
  pages={15--50},
  year={1996},
  publisher={Elsevier}
}

@article{kresse1993ab,
  title={Ab initio molecular dynamics for liquid metals},
  author={Kresse, Georg and Hafner, J{\"u}rgen},
  journal={Phys. Rev. B},
  volume={47},
  number={1},
  pages={558},
  year={1993},
  publisher={APS}
}

@article{kresse1999ultrasoft,
  title={From ultrasoft pseudopotentials to the projector augmented-wave method},
  author={Kresse, Georg and Joubert, Daniel},
  journal={Phys. Rev. B},
  volume={59},
  number={3},
  pages={1758},
  year={1999},
  publisher={APS}
}

@article{blochl1994projector,
  title={Projector augmented-wave method},
  author={Bl{\"o}chl, Peter E},
  journal={Phys. Rev. B},
  volume={50},
  number={24},
  pages={17953},
  year={1994},
  publisher={APS}
}

@article{phonopy-phono3py-JPCM,
  author  = {Togo, Atsushi and Chaput, Laurent and Tadano, Terumasa and Tanaka, Isao},
  title   = {Implementation strategies in phonopy and phono3py},
  journal = {J. Phys. Condens. Matter},
  volume  = {35},
  number  = {35},
  pages   = {353001},
  year    = {2023},
  doi     = {10.1088/1361-648X/acd831}
}

@article{phonopy-phono3py-JPSJ,
  author  = {Togo, Atsushi},
  title   = {First-principles Phonon Calculations with Phonopy and Phono3py},
  journal = {J. Phys. Soc. Jpn.},
  volume  = {92},
  number  = {1},
  pages   = {012001},
  year    = {2023},
  doi     = {10.7566/JPSJ.92.012001}
}

@article{kresse1995ab,
  title={Ab initio force constant approach to phonon dispersion relations of diamond and graphite},
  author={Kresse, G and Furthm{\"u}ller, J and Hafner, J},
  journal={Europhys. Lett.},
  volume={32},
  number={9},
  pages={729},
  year={1995},
  publisher={IOP Publishing}
}

@article{parlinski1997first,
  title={First-principles determination of the soft mode in cubic ZrO 2},
  author={Parlinski, K and Li, ZQ and Kawazoe, Y},
  journal={Phys. Rev. Lett.},
  volume={78},
  number={21},
  pages={4063},
  year={1997},
  publisher={APS}
}

@article{grimme2010consistent,
  title={A consistent and accurate ab initio parametrization of density functional dispersion correction (DFT-D) for the 94 elements H-Pu},
  author={Grimme, Stefan and Antony, Jens and Ehrlich, Stephan and Krieg, Helge},
  journal={J. Chem. Phys.},
  volume={132},
  number={15},
  year={2010},
  publisher={AIP Publishing}
}

@article{wood1985new,
  title={A new method for diagonalising large matrices},
  author={Wood, DM and Zunger, Alex},
  journal={J. Phys. A Math. Gen.},
  volume={18},
  number={9},
  pages={1343},
  year={1985},
  publisher={IOP Publishing}
}

@article{pulay1980convergence,
  title={Convergence acceleration of iterative sequences. The case of SCF iteration},
  author={Pulay, P{\'e}ter},
  journal={Chem. Phys. Lett.},
  volume={73},
  number={2},
  pages={393--398},
  year={1980},
  publisher={Elsevier}
}

@article{kim2024thermal,
  title={Thermal Hall effects due to topological spin fluctuations in {YMnO$_3$}},
  author={Kim, Ha-Leem and Saito, Takuma and Yang, Heejun and Ishizuka, Hiroaki and Coak, Matthew John and Lee, Jun Han and Sim, Hasung and Oh, Yoon Seok and Nagaosa, Naoto and Park, Je-Geun},
  journal={Nat. Commun.},
  volume={15},
  number={1},
  pages={243},
  year={2024},
  publisher={Nature Publishing Group UK London}
}

@article{zhang2021anomalous,
  title={Anomalous Thermal Hall Effect in an Insulating van der Waals Magnet},
  author={Zhang, Heda and Xu, Chunqiang and Carnahan, Caitlin and Sretenovic, Milos and Suri, Nishchay and Xiao, Di and Ke, Xianglin},
  journal={Phys. Rev. Lett.},
  volume={127},
  number={24},
  pages={247202},
  year={2021},
  publisher={APS}
}

@article{xu2024thermal,
  title={Thermal Hall effect in the van der Waals ferromagnet {CrI$_3$}},
  author={Xu, Chunqiang and Zhang, Heda and Carnahan, Caitlin and Zhang, Pengpeng and Xiao, Di and Ke, Xianglin},
  journal={Phys. Rev. B},
  volume={109},
  number={9},
  pages={094415},
  year={2024},
  publisher={APS}
}

@article{xu2023thermal,
  title={Thermal Hall effect in a van der Waals triangular magnet {FeCl$_2$}},
  author={Xu, Chunqiang and Carnahan, Caitlin and Zhang, Heda and Sretenovic, Milos and Zhang, Pengpeng and Xiao, Di and Ke, Xianglin},
  journal={Phys. Rev. B},
  volume={107},
  number={6},
  pages={L060404},
  year={2023},
  publisher={APS}
}

@article{bao2023direct,
  title={Direct observation of topological magnon polarons in a multiferroic material},
  author={Bao, Song and Gu, Zhao-Long and Shangguan, Yanyan and Huang, Zhentao and Liao, Junbo and Zhao, Xiaoxue and Zhang, Bo and Dong, Zhao-Yang and Wang, Wei and Kajimoto, Ryoichi and others},
  journal={Nat. Commun.},
  volume={14},
  number={1},
  pages={6093},
  year={2023},
  publisher={Nature Publishing Group UK London}
}

@article{ideue2017giant,
  title={Giant thermal Hall effect in multiferroics},
  author={Ideue, T and Kurumaji, T and Ishiwata, S and Tokura, Y},
  journal={Nat. Mater.},
  volume={16},
  number={8},
  pages={797--802},
  year={2017},
  publisher={Nature Publishing Group UK London}
}

@article{liechtenstein1995density,
  title={Density-functional theory and strong interactions: Orbital ordering in Mott-Hubbard insulators},
  author={Liechtenstein, AI and Anisimov, Vladimir I and Zaanen, Jan},
  journal={Phys. Rev. B},
  volume={52},
  number={8},
  pages={R5467},
  year={1995},
  publisher={APS}
}

@article{lebert2022acoustic,
  title={Acoustic phonon dispersion of {$\alpha$-RuCl$_3$}},
  author={Lebert, Blair W and Kim, Subin and Prishchenko, Danil A and Tsirlin, Alexander A and Said, Ayman H and Alatas, Ahmet and Kim, Young-June},
  journal={Phys. Rev. B},
  volume={106},
  number={4},
  pages={L041102},
  year={2022},
  publisher={APS}
}

\clearpage

\begin{widetext}

\section{}

\subsection{Supplemental Note 1. Phonon Hall Viscosity as Electronic Nuclear Berry Curvature}
In this first supplemental section, we discuss the connection between the phonon Hall viscosity and nuclear Berry curvature from the perspective of an effective phonon action.  The total action for a coupled phonon and electronic system can be written:
\begin{align}
S \equiv \int dt \ \mathcal{L} =  \int dt \  \langle \Psi | i\hbar \frac{\partial}{\partial t} - \mathcal{H}|\Psi\rangle
\end{align}
where $\mathcal{L}$ is the Lagrangian and $|\Psi\rangle$ is the full wavefunction including both electronic and lattice degrees of freedom. The trajectory of a vibrational wavepacket is determined by the stationary point of the phase $iS/\hbar = \gamma_{\rm geo} + \gamma_{\rm dyn}$. The first contribution $\gamma_{\rm geo} = - \int dt \ \langle \Psi | \frac{\partial}{\partial t}|\Psi\rangle$ is the geometric phase, while the second contribution $\gamma_{\rm dyn} = (i/\hbar)\int dt \ \langle \Psi | \mathcal{H}|\Psi\rangle$ is the dynamical phase. We consider the adiabatic approximation for the total wavefunction $|\Psi\rangle = |\Psi_{\rm ph}\rangle |\Psi_{\rm el}[\mathbf{u}(t)]\rangle$, where $|\Psi_{\rm ph}\rangle$ denotes the phonon (lattice) wavefunction, $|\Psi_{\rm el}\rangle$ denotes the electronic wavefunction, and $\mathbf{u}(t)$ are the time-dependent displacements of the atoms from equilibrium. 
If the dynamics of the lattice are slow compared to the electronic system, then $|\Psi_{\rm el}\rangle$ remains in the ground state. However, the explicit dependence of the ground state $|\Psi_{\rm el}\rangle$ on $\mathbf{u}(t)$ gives an additional contribution to the geometric phase $\gamma_{\rm el-ph} = -\int dt \ \langle \Psi_{\rm el} |  \nabla_{\mathbf{u}}\Psi_{\rm el}\rangle \cdot \frac{\partial \mathbf{u}}{\partial t}$. Putting everything together leads to:
\begin{align}
\mathcal{L} = \sum_{q\nu} \frac{1}{2}m_{q\nu} \dot{u}_{q\nu}^\dagger \dot{u}_{q\nu} 
-  \frac{1}{2} m_{q\nu} \omega_{q\nu}^2 u_{q\nu}^\dagger u_{q\nu}
+ \langle \Psi_{\rm el} |  i\hbar \frac{\partial}{\partial u_{q\nu}}|\Psi_{\rm el}\rangle \ \dot{u}_{q\nu} 
- \langle \Psi_{\rm el} |\mathcal{H}_{\rm el} [ \mathbf{u}(t)]|\Psi_{\rm el}\rangle
\end{align}
where $\dot{u} = \partial u / \partial t$ and $u_{q\nu}^\dagger = u_{-q\nu}$. The summation is over wavevectors $q$ and phonon band indices $\nu$. The third term represents a modification to the effective phonon Lagrangian. The essence of this result is that vibrational wavepackets are ``dressed'' by the electronic system, and can acquire an additional geometric phase due to the evolution of the ``dressing''. The quantity $i \langle \Psi_{\rm el} |  \nabla_{\mathbf{u}}\Psi_{\rm el}\rangle $ 
plays an analogous role to the magnetic vector potential for charged particles.
The specific details of this emergent gauge field are determined by the details of electron-phonon coupling and the electronic ground state. If the electronic ground state breaks time-reversal symmetry (either spontaneously or in response to an external magnetic field), this can lead to a finite phonon THE \cite{saito2019berry}.

The nuclear Berry connection can be evaluated perturbatively in powers of the electron-phonon coupling $\sum_{q\nu} \mathcal{O}_{q\nu} u_{q\nu}$. Taking $|g\rangle$ to be the unperturbed electronic ground state (in the absence of spin-phonon coupling), and $|m\rangle$ to be unperturbed excited states, then:
\begin{align}
|\Psi_{\rm el}\rangle \approx |g\rangle + \sum_{q\nu} \sum_m |m\rangle \frac{\langle m | \mathcal{O}_{q\nu} |g\rangle}{E_g-E_m}u_{q\nu} 
\end{align}
This leads to:
\begin{align}
i\hbar \left\langle \Psi_{\rm el} \left|\frac{\partial \Psi_{\rm el}}{\partial u_{q\nu}}\right.\right\rangle \approx i\hbar  \sum_{\nu^\prime,m} u_{q\nu^\prime}^\dagger\frac{\langle g|\mathcal{O}_{q\nu^\prime}^\dagger |m\rangle\langle m | \mathcal{O}_{q\nu}|g\rangle}{(E_g-E_m)^2}  = i\hbar \sum_{\nu^\prime} \left\langle \frac{\partial\Psi_{\rm el}}{\partial u_{q\nu^\prime}^\dagger} \left|\frac{\partial \Psi_{\rm el}}{\partial u_{q\nu}}\right.\right\rangle u_{q\nu^\prime}^\dagger
\\
i\hbar \left\langle\left. \frac{\partial \Psi_{\rm el}}{\partial u_{q\nu}} \right|\Psi_{\rm el} \right\rangle \approx i\hbar \sum_{\nu^\prime,m} \frac{\langle g|\mathcal{O}_{q\nu} |m\rangle\langle m | \mathcal{O}_{q\nu^\prime}^\dagger|g\rangle}{(E_g-E_m)^2} u_{q\nu^\prime}^\dagger = i\hbar \sum_{\nu^\prime} \left\langle \frac{\partial\Psi_{\rm el}}{\partial u_{q\nu}} \left|\frac{\partial \Psi_{\rm el}}{\partial u_{q\nu^\prime}^\dagger}\right.\right\rangle u_{q\nu^\prime}^\dagger
\end{align}
One may note that normalization of the electronic wavefunction for all values of displacements requires that $\frac{\partial}{\partial u} \langle \Psi_{\rm el} | \Psi_{\rm el}\rangle =  \langle \frac{\partial \Psi_{\rm el} }{\partial u}| \Psi_{\rm el}\rangle +  \langle \Psi_{\rm el} |\frac{\partial \Psi_{\rm el} }{\partial u}\rangle = 0$, such that  $\langle \frac{\partial \Psi_{\rm el} }{\partial u}| \Psi_{\rm el}\rangle = -  \langle \Psi_{\rm el} |\frac{\partial \Psi_{\rm el} }{\partial u}\rangle$. As a consequence:
\begin{align}
i\hbar \left\langle \Psi_{\rm el} \left|\frac{\partial \Psi_{\rm el}}{\partial u_{q\nu}}\right.\right\rangle \approx \frac{i\hbar}{2} \sum_{\nu^\prime}\left(\left\langle \frac{\partial\Psi_{\rm el}}{\partial u_{q\nu^\prime}^\dagger}\left|\frac{\partial \Psi_{\rm el}}{\partial u_{q\nu}}\right.\right\rangle  - \left\langle \frac{\partial\Psi_{\rm el}}{\partial u_{q\nu}} \left|\frac{\partial \Psi_{\rm el}}{\partial u_{q\nu^\prime}^\dagger}\right.\right\rangle\right) u_{q\nu^\prime}^\dagger = \frac{\hbar}{2} \sum_{\nu^\prime} \Omega_q^{\nu^\prime\nu} u_{q\nu^\prime}^\dagger
\end{align}
Here, we have introduced the nuclear Berry curvature in $q$-space:
\begin{align}\label{eq:nuclearBC}
 \Omega_q^{\nu^\prime \nu} = 
i \left(\left\langle \frac{\partial\Psi_{\rm el}}{\partial u_{q\nu^\prime}^\dagger}\left|\frac{\partial \Psi_{\rm el}}{\partial u_{q\nu}}\right.\right\rangle  - \left\langle \frac{\partial\Psi_{\rm el}}{\partial u_{q\nu}} \left|\frac{\partial \Psi_{\rm el}}{\partial u_{q\nu^\prime}^\dagger}\right.\right\rangle\right) = i\sum_{m}\frac{\langle g|\mathcal{O}_{q\nu^\prime}^\dagger |m\rangle\langle m | \mathcal{O}_{q\nu}|g\rangle-\langle g|\mathcal{O}_{q\nu} |m\rangle\langle m | \mathcal{O}_{q\nu^\prime}^\dagger|g\rangle}{(E_g-E_m)^2}
\end{align}
For reasons made clear below, it is convenient to absorb the prefactors of the Fourier transformed phonon operators, and represent the nuclear Berry curvature in terms of the reduced Hall viscosity $\overline{\eta}_q^{\nu^\prime \nu} = -\overline{\eta}_{-q}^{\nu\nu^\prime}$:
\begin{align}
\overline{\eta}_q^{\nu^\prime \nu} = & \ \sqrt{\frac{\hbar}{2m_{q\nu}\omega_{q\nu}}} \sqrt{\frac{\hbar}{2m_{q\nu^\prime} \omega_{q\nu^\prime}}}  \ \Omega_q^{\nu^\prime \nu}
\end{align}
such that:
\begin{align}
i\hbar\left\langle \Psi_{\rm el} \left|\frac{\partial \Psi_{\rm el}}{\partial u_{q\nu}}\right.\right\rangle \approx & \ \sum_{\nu^\prime} \sqrt{m_{q\nu}\omega_{q\nu}m_{q\nu^\prime}\omega_{q\nu^\prime}} \ \overline{\eta}_{q}^{\nu^\prime \nu}  \ u_{q\nu^\prime}^\dagger 
\end{align}
 The above expressions are analogous to the symmetric Coulomb gauge for the electromagnetic vector potential, wherein $\vec{A} = -\frac{1}{2} \vec{r}\times \vec{B}$, except that $ \Omega_q^{\nu^\prime \nu} $ plays the role of the magnetic field. As defined, $\Omega$ has units of inverse length squared, and $\overline{\eta}$ is unitless. The relationship between this quantity and the usual definition of the viscosity in the long-wavelength elastic theory is discussion in Supplementary Section S4.

Returning to the effective phonon Lagrangian, we find:
\begin{align}
\mathcal{L}_{\rm eff} =\sum_{q\nu} \frac{1}{2}m_{q\nu} \dot{u}_{q\nu}^\dagger \dot{u}_{q\nu} 
-  \frac{1}{2} m_{q\nu} \omega_{q\nu}^2 u_{q\nu}^\dagger u_{q\nu}
+\sum_{q\nu \nu^\prime} \sqrt{m_{q\nu}\omega_{q\nu}m_{q\nu^\prime}\omega_{q\nu^\prime}} \  u_{q\nu^\prime}^\dagger \overline{\eta}_{q}^{\nu^\prime \nu} \dot{u}_{q\nu}
\label{eq:lagrange}
\end{align}
From this, we define the phonon momenta:
\begin{align}
p_{q\nu}^\dagger = \frac{\partial \mathcal{L}_{\rm eff}}{\partial \dot{u}_{q\nu}} = m_{q\nu} \dot{u}_{q\nu}^\dagger +\sum_{\nu^\prime} \sqrt{m_{q\nu}\omega_{q\nu}m_{q\nu^\prime}\omega_{q\nu^\prime}} \  u_{q\nu^\prime}^\dagger \overline{\eta}_{q}^{\nu^\prime \nu} 
\end{align}
The resulting effective phonon Hamiltonian up to linear order in $\eta$ is:
\begin{align}
\mathcal{H}_{\rm eff} 
= & \ \sum_{q\nu} \frac{p_{q\nu}^\dagger p_{q\nu}}{2m_{q\nu}} 
+\frac{1}{2} m_{q\nu} \omega_{q\nu}^2 u_{q\nu}^\dagger u_{q\nu}
-\sum_{q\nu\nu^\prime} \frac{\sqrt{m_{q\nu}\omega_{q\nu}m_{q\nu^\prime}\omega_{q\nu^\prime}} }{2}\left(  u_{q\nu}^\dagger \overline{\eta}_q^{\nu \nu^\prime}\frac{p_{q\nu^\prime}}{m_{q\nu^\prime}}-\frac{p_{q\nu}^\dagger}{m_{q\nu}} \overline{\eta}_{q}^{\nu \nu^\prime }u_{q\nu^\prime} \right) \label{eq:hefffull}
\end{align}
We see that the nuclear Berry curvature couples to an antisymmetric combination of the phonon displacement and momentum resembling a phonon angular momentum. 

As discussed in Ref.~\onlinecite{ye2020phonon,ye2021phonon}, the Hall viscosity can be separately identified with the antisymmetric part of the adiabatic limit of the phonon self-energy. At one loop order in the spin-phonon coupling, the self-energy is defined by:
\begin{align}
\Pi_q^{\nu\nu^\prime} (i\omega_n)= & \ -\frac{1}{2\hbar}\int_{-\hbar\beta}^{\hbar\beta} d\tau \  e^{i\omega_n \tau}\langle\mathcal{T}_\tau [\mathcal{O}_{q\nu}(\tau) \mathcal{O}_{q\nu^\prime}^\dagger (0)] \rangle
\\
= & \   \frac{1}{2}\sum_{nm} \frac{e^{-\beta E_n}-e^{-\beta E_m} }{Z_{\rm el}} \left[  \frac{\langle n|  \mathcal{O}_{q\nu^\prime}^\dagger |m\rangle \langle m | \mathcal{O}_{q\nu}|n \rangle}{ E_n-E_m-\hbar \ i\omega_n}+
        \frac{\langle n|  \mathcal{O}_{q\nu} |m\rangle\langle m| \mathcal{O}_{q\nu^\prime}^\dagger |n \rangle}{E_n-E_m+\hbar \ i\omega_n} 
        \right] 
\end{align}
where $|n\rangle$ and $|m\rangle$ are unperturbed electronic states and $Z_{\rm el}$ is the unperturbed electronic partition function. The reduced Hall viscosity is then given by:
\begin{align}
\overline{\eta}_q^{\nu^\prime \nu } =  & \ \sqrt{\frac{1}{2m_{q\nu}\omega_{q\nu}}}\sqrt{\frac{1}{2m_{q\nu^\prime}\omega_{q\nu^\prime}}} \lim_{i\omega_n \to 0}     \frac{i}{i\omega_n}\Pi_q^{\nu\nu^\prime}
\\
= & \ \frac{i}{2} \sqrt{\frac{\hbar}{2m_{q\nu}\omega_{q\nu}}}\sqrt{\frac{\hbar}{2m_{q\nu^\prime}\omega_{q\nu^\prime}}}\sum_{nm} \frac{e^{-\beta E_n}-e^{-\beta E_m} }{Z_{\rm el}} \left[ 
  \frac{\langle n|  \mathcal{O}_{q\nu^\prime}^\dagger |m\rangle \langle m | \mathcal{O}_{q\nu}|n \rangle
  -\langle n|  \mathcal{O}_{q\nu} |m\rangle\langle m| \mathcal{O}_{q\nu^\prime}^\dagger |n \rangle}{ (E_n-E_m)^2}\right]
\end{align}
which is simply the finite-temperature analogue of the nuclear Berry curvature defined above.

\subsection{Supplemental Note 2. Details of First Principles Calculations}
In this section, we describe the full computational details of the spin-phonon couplings. There are four essential steps. 
\begin{enumerate}
\item The structure is relaxed to obtain a starting geometry for further calculations.
\item \textit{Ab-initio} phonon calculations are performed on the relaxed structure to obtain phonon frequencies and eigenvectors. A gauge transformation is applied to the eigenvectors to ensure a smooth definition at finite $|q|$. 
\item Linear electron-phonon couplings are computed on the relaxed structure for a Wannier basis of $d$-orbitals on each Ru site using a finite displacement approach. 
\item Exact diagonalization of the $d$-orbital electron+phonon Hamiltonian including electronic interactions and electron-phonon coupling is performed for small clusters of Ru atoms. The resulting low-energy states are projected onto pure $j_{1/2}$ states with variable numbers of phonon quanta, and the low-energy Hamiltonian is analyzed to extract generic spin-phonon couplings. 
\end{enumerate}

{\it Structural Relaxation:} We first relaxed an idealized structure of \rucl with AA stacking in the space group P$\bar{3}$1m, starting from a similar out-of-plane layer-to-layer distance as observed in the reported $R\bar{3}$ structure. For this purpose, the Vienna \textit{Ab Initio} Simulation Package (VASP) \cite{kresse1996efficient,kresse1996efficiency,kresse1993ab} was used for the DFT calculations with PBE-GGA \cite{perdew1996generalized} exchange correlation functional. The Projector augmented wave (PAW) pseudopotential  \cite{kresse1999ultrasoft,blochl1994projector} was implemented for the given elements as provided in VASP. To account the van der Waals interaction, we used the DFT-D3 method of Grimme with zero-damping function \cite{grimme2010consistent} and for the electronic minimization algorithm, we choose a mixture of the blocked-Davidson and RMM-DIIS algorithms \cite{wood1985new,pulay1980convergence}. To approximately treat the strongly correlated \textit{d}-electrons, we employed the rotationaly invariant DFT+U of  Liechtenstein \textit{et al.} \cite{liechtenstein1995density} with U = 1.80 eV and J = 0.40 eV on \textit{d} orbitals of Ru atoms (method adapted from \cite{lebert2022acoustic}). The relaxation was performed with an initial antiferromagnetic (Ne\'el) spin configuration on the Ru sites. A k-mesh sampling of $8 \times 8 \times 8$  was chosen for the first Brillouin zone and the structural relaxation was performed until the forces were less than 0.005 eV/\AA. The resulting structure is depicted in Fig.~\ref{fig:structure}, and has a c-axis lattice constant of 5.7389 \AA \  and in-plane lattice constant of 6.0577 \AA.

\begin{figure}[b]
\includegraphics[width=0.8\linewidth]{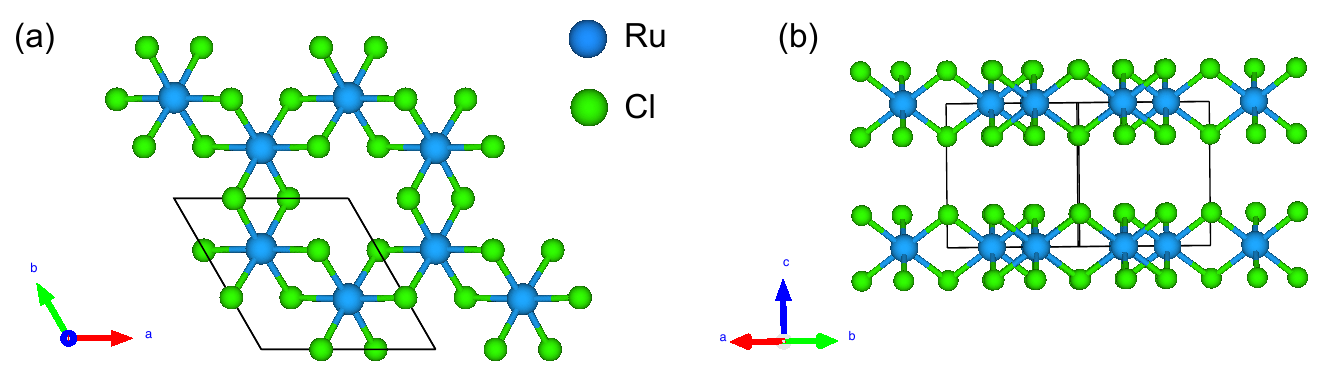}
\caption{Relaxed high-symmetry crystal structure of \rucl with the P$\bar{3}$1m symmetry (a): view along $c$-axis showing honeycomb planes. (b): side view showing AA stacking. Ru and Cl atoms are shown in blue and green colors respectively.}  
\label{fig:structure}
\end{figure}

{\it Ab-initio Phonons:} To carry out the phonon calculations, we prepared a super-cell of $2 \times 2 \times 2$ and utilized the finite displacement method \cite{kresse1995ab,parlinski1997first} as implemented in Phonopy \cite{phonopy-phono3py-JPCM,phonopy-phono3py-JPSJ}. Here, we chose a less dense k-mesh of $4 \times 4 \times 4$ due to the reduced Brillouin zone size of supercell.  From these calculations, we obtain the phonon eigenvectors $e_{\alpha n}^{q\nu}$ and eigenenergies $\hbar \omega_{q\nu}$. In terms of $q-$space operators, the real-space atomic displacement and momentum operators may be written:
\begin{align}\label{eqn:uln}
\hat{u}_\alpha (\ell n) = & \ \sqrt{\frac{\hbar}{2Nm_n}} \sum_{q\nu} \frac{e^{iq\cdot r_{\ell n}}}{\sqrt{\omega_{q\nu}} }\ (a_{-q\nu}^\dagger + a_{q\nu}  ) \ e_{\alpha n}^{q\nu} 
=  \frac{1}{\sqrt{N}} \sum_{q\nu} e^{iq\cdot r_{\ell n}} \sqrt{\frac{m_{q\nu}}{m_n}} \ e_{\alpha n}^{q\nu} \ \hat{u}_{q\nu}
\\ \label{eqn:pln}
\hat{p}_\alpha (\ell n) =& \ i \sqrt{\frac{\hbar m_n}{2N} }\sum_{q\nu} \sqrt{\omega_{q\nu} }\ e^{iq\cdot r_{\ell n}} (a^\dagger_{-q\nu} - a_{q\nu}) \ e_{\alpha n}^{q\nu}
=\frac{1}{\sqrt{N} }\sum_{q\nu} e^{iq\cdot r_{\ell n}} \sqrt{\frac{m_n}{m_{q\nu}}} \ e_{\alpha n}^{q\nu} \  \hat{p}_{q\nu}
\end{align}
where $\hat{u}_\alpha (\ell n)$ refers to the displacement of the $n$th atom in the $\ell$th unit cell, in the $\alpha\in\{x,y,z\}$ direction. $r_{\ell n}$ is the position of such atom, $m_n$ is it's mass, and $e_{\alpha n}^{q\nu} = (e_{\alpha n}^{-q\nu})^*$ are the phonon eigenvectors associated with momentum $q$ and band index $\nu$.

{\it Phonon Gauge Choice:} For the acoustic modes, we then make a gauge transformation to ensure a smooth definition of $e_{\alpha n}^{q\nu}$ at finite in-plane $|q|$. The gauge transformation is equivalent to making a $q,\nu$-dependent shift of the origin of the Fourier transform, which results in:
\begin{align}
 e_{\alpha n}^{q\nu}\to & \ e_{\alpha n}^{q\nu} e^{-i\phi_{q\nu}}
\\
\mathcal{A}_{q\nu}\to & \ \mathcal{A}_{q\nu}e^{i\phi_{q\nu}}
\\
\mathcal{L}_{q\nu;q^\prime \nu^\prime}\to & \ \mathcal{L}_{q\nu;q^\prime \nu^\prime} e^{i(\phi_{q\nu}+\phi_{q^\prime\nu^\prime})}
\end{align}
which retains the feature that $e_{\alpha n}^{q\nu} = (e_{\alpha n}^{-q\nu})^*$ provided $\phi_{q\nu} = -\phi_{-q\nu}$. As depicted in Fig.~\ref{fig:phonon_gauge}, it is not possible for both the real and imaginary parts of the phonon eigenvectors to be continuous at both $q=0$ and finite $q$ if $e_{\alpha n}^{q\nu} = (e_{\alpha n}^{-q\nu})^*$ is satisfied. For example, there is an obstruction on $\text{Re}[e_{\alpha n}^{q\nu}]$ for the LA and TA modes that can be seen from the requirement that the primary direction of atomic displacement should rotate with the $q$-vector if it is to remain parallel or transverse to $q$. A smooth gauge therefore requires $e_{\alpha n}^{q\nu} = -e_{\alpha n}^{-q\nu}$ which  excludes a real component to the eigenvector since $\text{Re}[e_{\alpha n}^{q\nu}] =  \text{Re}[e_{\alpha n}^{-q\nu}]$. In the conventional gauge choice employed in most {\it ab-initio} phonon codes, the eigenvectors are defined such that $\text{Im}[e_{\alpha n}^{q\nu}]$ vanishes in the limit $q\to 0$. In this case, it is necessary for the eigenvectors of the TA and LA modes to have branch cuts in $\text{Re}[e_{\alpha n}^{q\nu}]$ at finite $q$, leading to corresponding branch cuts in the spin-phonon couplings. For these modes, it is more convenient to make a gauge transformation that makes the eigenvectors continuous at all finite $q$. The eigenvectors of the TA and LA modes can be made continuous everywhere in the 2D Brillouin zone except exactly at $q=0$ via a transformation that makes $e_{\alpha n}^{q\nu}$ primarily imaginary, such that $\lim_{q\to 0}\text{Re}[e_{\alpha n}^{q\nu}] = 0$. Similarly, the eigenvectors for the ZA modes are made continuous by ensuring that the imaginary component vanishes in the small momentum limit $\lim_{q\to 0}\text{Im}[e_{\alpha n}^{q\nu}] = 0$.
The consequences for the spin-phonon couplings is discussed below.

\begin{figure}[t]
\includegraphics[width=0.4\linewidth]{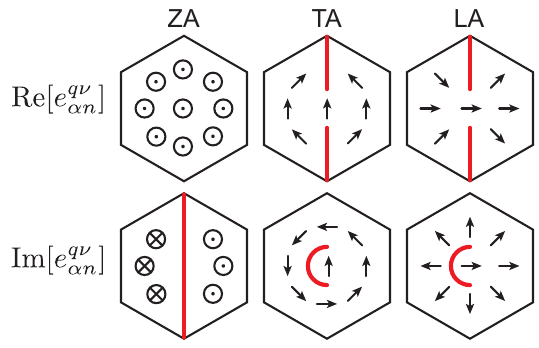}
\caption{Evolution of the real and imaginary parts of the phonon eigenvectors $e_{\alpha n}^{q\nu}$ for ZA, TA, and LA modes in the chosen gauge. Arrows indicate the dominant direction of atomic displacement corresponding to eigenvectors at different $q$-points in the Brillouin zone. Transforming the ZA eigenvectors to be purely real and the TA and LA eigenvectors to be purely imaginary in the limit $q\to 0$ ensures smooth spin-phonon couplings in the same limit.}
\label{fig:phonon_gauge}
\end{figure}

{\it Wannier Hamiltonian:} In order to estimate the spin-phonon couplings, we first estimate the one-particle contributions to the Hamiltonian in a basis of $d$-orbital Wannier functions on each Ru site. Formally, the one-particle Hamiltonian includes intersite hopping, intrasite crystal field, and spin-orbit coupling, $\mathcal{H}_{1p} = \mathcal{H}_{hop}+\mathcal{H}_{\rm CF} + \mathcal{H}_{\rm SO}$:
\begin{align}
    \mathcal{H}_{hop} =& \  \sum_{ij\alpha\beta\sigma}t_{ij}^{\alpha\beta} c_{i,\alpha,\sigma}^\dagger c_{j,\beta,\sigma}
    \\
    \mathcal{H}_{\rm CF} = & \ \sum_{i\alpha\beta\sigma}d_{i}^{\alpha\beta}c_{i,\alpha,\sigma}^\dagger c_{i,\beta,\sigma}
    \\
    \mathcal{H}_{\rm SO} = & \ \sum_{i\alpha\beta\sigma\sigma^\prime} \lambda_{\rm Ru} \langle \phi_i^\alpha(\sigma)|\mathbf{L}\cdot\mathbf{S}|\phi_{i}^\beta(\sigma^\prime)\rangle c_{i,\alpha,\sigma}^\dagger c_{i,\beta,\sigma^\prime}
\end{align}
where $c_{i,\alpha,\sigma}^\dagger$ creates an electron at Ru site $i$, in $d$-orbital $\alpha$, with spin $\sigma$. Here, $\mathbf{L}$ is the orbital momentum operator (not to be confused with the spin-phonon coupling addressed below). The linear electron-phonon coupling represents the modulation of these terms that is linear in atomic displacements: 
\begin{align}\label{eq:elphon}
\mathcal{H}_{\rm el-ph} = & \ \sum_{\alpha \ell n} \mathcal{H}_{\Delta}^{\alpha \ell n} \hat{u}_\alpha (\ell n)
\\ 
\mathcal{H}_{\rm \Delta}^{\alpha \ell n} =& \  \sum_{i\sigma j\sigma^\prime}\Delta_{i\sigma j\sigma^\prime}^{\mu\nu;\ell\alpha n}\  c_{i,\mu,\sigma}^\dagger c_{j,\nu,\sigma^\prime}
\end{align}
The phonon displacement operators $\hat{u}_\alpha (\ell n)$ are defined above in Eq.~(\ref{eqn:uln}) and refer to displacement of atom $n$, in the unit cell labelled $\ell$, in the direction $\alpha \in \{x,y,z\}$. The elements $\Delta_{i\sigma j\sigma^\prime}^{\mu\nu;\ell\alpha n}$ refer to changes in the single-particle matrix elements due to such a displacement, including all the hopping, CF, and SOC terms.

In order to estimate all of the undisplaced one-particle terms, we perform fully relativistic density functional theory calculations on a 2$\times$2$\times$1 supercell of the relaxed $P\bar{3}1m$ structure using FPLO \cite{opahle1999full}, and project \cite{koepernik2023symmetry} the resulting Kohn-Sham bands onto Ru $d$-orbitals to obtain the electronic Hamiltonian in terms of Ru Wannier functions. For this purpose, we employ the PBE functional\cite{perdew1996generalized}, and a 6$\times$6$\times$1 $k$-point mesh. 

To estimate the electron-phonon coupling (in real-space), we take a supercell approach similar to that described in Ref.~\onlinecite{togo2023first}. Using the same 2$\times$2$\times$1 supercell, we compute the changes in the single-particle terms between sites in the full supercell induced by small displacements of each atom in the parent primitive cell. For this purpose, we displace each atom along the cartesian $x,y,z$ directions by $\pm$0.02 \AA, and perform a DFT calculation on each geometry using FPLO and Wannier projection. The changes in the hoppings are then obtained by taking finite differences of the displaced and undisplaced Hamiltonians in the Wannier basis. Finally, each orbital in the supercell is translated back to the parent cell to identify the corresponding $\Delta_{i\sigma j\sigma^\prime}^{\mu\nu;\ell\alpha n}$ term with respect to the primitive cell. This provides an estimate of the electron-phonon coupling associated with any pair of orbitals, and an atomic displacement within a supercell distance of one of those orbitals. This coupling falls off rapidly with distance between the orbital site and the displaced atom, such that a 2$\times$2$\times$1 supercell suffices to capture the relevant effects within the monolayer.

{\it Spin-Phonon Couplings:} Finally, we employ the des Cloizeaux effective Hamiltonian (dCEH) approach outlined in Ref.~\cite{dhakal2024mn} to obtain the spin-phonon couplings. As noted in the main text, the spin-phonon Hamiltonian is defined as:
\begin{align}
    \mathcal{H}_{\rm sp-ph} = &  \sum_{q\nu} \bar{u}_{q\nu} \  \mathcal{A}_{q\nu}+\sum_{q\nu q^\prime \nu^\prime } \bar{u}_{q\nu} \bar{p}_{q^\prime \nu^\prime} \  \mathcal{L}_{q\nu ;q^\prime \nu^\prime}
\end{align}
where $\bar{u}_{q\nu} = (a_{-q\nu}^\dagger + a_{q\nu})$ and $\bar{p}_{q\nu} = i(a_{-q\nu}^\dagger - a_{q\nu})$, and $a_{q\nu}^\dagger$ creates a phonon with momentum $q$ in band $\nu$. Here, the units for the phonon operators have been absorbed into the $\mathcal{A}$ and $\mathcal{L}$ operators, so that the latter have units of energy. The $q$-space spin and bond operators are:
\begin{align}
\label{eqn:Adefsup}
\mathcal{A}_{qv} = & \ \frac{1}{\sqrt{N}}\sum_{ij} \left(\mathbf{S}_i \cdot \mathbb{A}_{ij}^{q\nu}\cdot \mathbf{S}_j\right) \ e^{-iq\cdot (r_i+r_j)/2}
\\ \label{eqn:Ldefsup}
\mathcal{L}_{q\nu ;q^\prime \nu^\prime} =& \  \frac{\hbar\omega_{q^\prime\nu^\prime}}{N}\sum_{i}\left(\mathbf{L}_{i}^{q\nu ;q^\prime \nu^\prime} \cdot \mathbf{S}_i\right)\ e^{-i(q+q^\prime)\cdot r_i}
\end{align}
where $N$ is the number of unit cells, and $\mathbf S_i$ describe the $j_{1/2}$ moments. To compute $\mathbf{L}$ and $\mathbb{A}$, we implemented a real-space and momentum-space formulation of this approach, which have different merits. For the $\mathcal{A}$ couplings, we emlpoyed the real-space formulation. For each bond of interest, and each elementary real-space displacement $(\alpha \ell n)$, we construct a local many-body model including two Ru sites. The Hamiltonian of the local models takes the form:
 \begin{align}
 \mathcal{H}_{ij}(\alpha\ell n)= \mathcal{H}_{1p}+\mathcal{H}_{2p}+\mathcal{H}_{\Delta}^{\alpha\ell n} \  (a^\dagger + a) \ \delta u + \hbar\omega_0 a^\dagger a
 \end{align}
where $\mathcal{H}_{1p}$ and $\mathcal{H}_{\Delta}^{\alpha\ell n}$ are obtained as described above. The operator $a^\dagger$ creates an auxiliary phonon, which is included as a degree of freedom corresponding to a displacement of $\delta u = 0.01$ \AA \ and energy $\hbar\omega_0 = 1$ meV.  The two-particle interactions are taken to be:
\begin{align}
\mathcal{H}_{2p} = \sum_{i\alpha\beta\delta\gamma}\sum_{\sigma\sigma^\prime}U_{\alpha\beta\gamma\delta} \ c_{i,\alpha,\sigma}^\dagger c_{i,\beta,\sigma^\prime}^\dagger c_{i,\gamma,\sigma^\prime} c_{i,\delta,\sigma}
\end{align}
where $U_{\alpha\beta\gamma\delta}$ are parameterized by the Slater parameters $F_0^{dd}, F_2^{dd}, F_4^{dd}$, following the spherically symmetric approximation \cite{sugano1970multiplets}. In terms of these parameters, the Kanamori parameters for the $t_{2g}$ orbitals satisfy: 
\begin{align}
U_{t2g} = & \ F_0^{dd} + \frac{4}{49}(F_2^{dd} + F_4^{dd}) 
\\
J_{t2g} = & \ \frac{3}{49}F_2^{dd} + \frac{20}{441}F_4^{dd}
\end{align}
We use $U_{t2g} = 2.58$ eV, $J_{t2g} = 0.29$ eV, and $F_4^{dd} = (5/8)F_2^{dd}$, which are taken from previous theoretical estimates from constrained RPA \cite{eichstaedt2019deriving}, and roughly compatible with estimates from optical studies \cite{sandilands2016spin,sandilands2016optical,lebert2023nonlocal}. The electron/phonon Hamiltonian $\mathcal{H}$ is then exactly diagonalized within a basis including $n = 0$ or 1 phonon quanta, and the resulting low-energy states projected onto ideal $j_{1/2}$ states with variable number of phonon quanta as described in Ref.~\cite{dhakal2024mn}. The low-energy Hamiltonian obtained takes the form:
\begin{align}
\mathcal{H}_{\rm low} = \left(\mathbf{S}_i \cdot \mathbb{J}_{ij} \cdot \mathbf{S}_j \right) + \left(\mathbf{S}_i \cdot \mathbb{A}_{ij}^{\alpha\ell n} \cdot \mathbf{S}_j \right)(a^\dagger + a) \ \delta u \ + i\hbar\omega_0 \left(\mathbf{G}_i^{\alpha\ell n} \cdot \mathbf{S}_i +\mathbf{G}_j^{\alpha\ell n} \cdot \mathbf{S}_j \right)(a^\dagger - a) \ \delta u  +\mathcal{O}(\delta u^2)
\end{align}
Here, $\mathbb{A}_{ij}^{\alpha\ell n} $ is the change in the intersite coupling matrix that is linear in atomic displacement. This is obtained by dividing the terms symmetric in $a^\dagger$ and $a$ by $\delta u$. It may be noted that the computed $\mathbb{A}_{ij}^{\alpha\ell n} $ matrices are insensitive to the specific choice of $\delta u$ provided it is sufficiently small to correspond to a linear perturbative regime. We confirmed linear scaling by checking different values of $\delta u$ in the range of 0.01 \AA to 0.05 \AA. $\mathbf{G}_i$ is a symmetry allowed coupling of the local spin to the phonon momentum, which is obtained by dividing the terms antisymmetric in $a^\dagger$ and $a$ by $i\hbar\omega_0 \delta u$. The origin and scaling of this term is discussed in Ref.~\cite{dhakal2024mn}. We compute it to be small $\lesssim 10^{-2}$ meV for all bands, and thus neglect it in our analysis. 

After repeating the calculation for all elementary displacements in real-space, the $q$-space operator is then constructed via:
\begin{align}
 \label{eq:Aqvsup}
    \mathbb{A}_{ij}^{qv} =& \  \sum_{\alpha \ell n} \sqrt{\frac{\hbar}{2m_n\omega_{qv}}}\  e_{\alpha n}^{qv} \  \mathbb{A}_{ij}^{\alpha\ell n} \ e^{iq\cdot r_{\ell n}}
\end{align}
This yields the full $q$- and band-dependence of the $ \mathbb{A}_{ij}^{qv}$ couplings. For a given bond, we included displacements of atoms within 4.0 \AA \ for the crystal field electron-phonon coupling, and 4.64 \AA \ of the bond center for the hopping electron-phonon terms. The electron-phonon couplings fall off rapidly with distance, so this truncation does not introduce significant errors. However, to mitigate small spurious effects associated with the finite distance cutoff, we subtract from each electron-phonon coupling (separately for each direction $\alpha$) the average value of $\Delta_{i\sigma j\sigma^\prime}^{\mu\nu;\ell\alpha n}$ over the atoms included within the cutoff. This ensures that any uniform translation of all atoms under consideration corresponds to no modification of the Hamiltonian (which ensures $\lim_{q\to 0} \mathbb{A} = 0$). The advantage of the real-space approach is that a relatively small number of exact diagonalization calculations (3 $\times$ number of bonds $\times$ number of atoms within cutoff) is required to obtain the full $q$-dependent spin-phonon couplings.

In order to estimate the two-phonon $\mathbf{L}$ couplings, we employed a different approach explicitly in $q$-space, which requires two ED calculations for each pair of phonon bands and $q$-point of interest. This formulation is less efficient unless one is interested in a small subset of bands or $q$-points. However, it has the advantage of mitigating numerical rounding errors accumulated over the summation of the real space spin-phonon couplings in the Fourier transform. For \rucl, the $\mathbf{L}$ couplings are orders of magnitude smaller than the $\mathbb{A}$ couplings, and require two Fourier transforms, making them more susceptible to rounding errors. We start by Fourier transforming the electron-phonon coupling  to $q$-space:
\begin{align}
\mathcal{H}_{\rm el-ph} = & \ \sum_{q\nu}   \mathcal{H}_{\Delta}^{q\nu}  \ u_{q\nu} = \sum_{q\nu} \mathcal{H}_{\Delta}^{q\nu} a_{q\nu} +  \mathcal{H}_{\Delta}^{-q\nu} a_{q\nu}^\dagger
\\
 \mathcal{H}_{\Delta}^{q\nu} = & \ \sum_{\alpha \ell n} 
\sqrt{\frac{\hbar}{2Nm_n}}\frac{e^{iq\cdot r_{\ell n}}}{\sqrt{\omega_{q\nu}} } \ e_{\alpha n}^{q\nu} \ \mathcal{H}_{\Delta}^{\alpha \ell n} 
\end{align}
where $\mathcal{H}_{\Delta}^{-q\nu} =  (\mathcal{H}_{\Delta}^{q\nu} )^\dagger$. We then consider one-site models for each pair of phonon bands:
\begin{align}
\mathcal{H}_i(q\nu q^\prime \nu^\prime) = & \ \mathcal{H}_{1p} + \mathcal{H}_{2p} +\left[ \mathcal{H}_{\Delta}^{-q\nu} a_{-q\nu} +  (\mathcal{H}_{\Delta}^{-q\nu} )^\dagger a_{-q\nu}^\dagger \right]
\delta u + \left[ \mathcal{H}_{\Delta}^{q^\prime \nu^\prime} a_{q^\prime \nu^\prime} +  (\mathcal{H}_{\Delta}^{q^\prime \nu^\prime} )^\dagger a_{q^\prime \nu^\prime}^\dagger \right]
\delta u^\prime 
\nonumber \\
& \ + \hbar \omega_0 \left( a_{q\nu}^\dagger a_{q\nu} +  a_{q^\prime \nu^\prime}^\dagger a_{q^\prime \nu^\prime}\right)
\end{align}
The electron/phonon Hamiltonian $\mathcal{H}_i$ is then exactly diagonalized within a basis including $n_{-q\nu},n_{q\nu^\prime} = 0$ or 1 phonon quanta, and the resulting low-energy states projected onto ideal $j_{1/2}$ states with variable number of phonon quanta. For $j_{1/2}$ moments, the resulting low-energy Hamiltonian is written:
\begin{align}
\mathcal{H}_{\rm low} = & \ 
i\hbar\omega_0\delta u\left(\mathbf{G}_i^{-q\nu}\cdot \mathbf{S}_i \right)  \left(a_{-q\nu}^\dagger - a_{-q\nu}\right)
+ i\hbar\omega_0^\prime \delta u^\prime \left( \mathbf{G}_i^{q\nu^\prime}   \cdot \mathbf{S}_i \right)  \left(a_{q\nu^\prime}^\dagger - a_{q\nu^\prime}\right)
\nonumber \\
& \ + \delta u \delta u^\prime \left( 
\mathbf{B}_{i;++}^{-q\nu;q^\prime\nu^\prime}  a_{-q\nu}^\dagger a_{q^\prime\nu^\prime}^\dagger 
+\mathbf{B}_{i;+-}^{-q\nu;q^\prime\nu^\prime}  a_{-q\nu}^\dagger a_{q^\prime\nu^\prime} 
+\mathbf{B}_{i;-+}^{-q\nu;q^\prime\nu^\prime}  a_{-q\nu} a_{q^\prime\nu^\prime}^\dagger 
+\mathbf{B}_{i;--}^{-q\nu;q^\prime\nu^\prime}  a_{-q\nu} a_{q^\prime\nu^\prime} 
\right)\cdot\mathbf{S}_i + ...
\end{align}
We then approximate the $\mathbf{L}$ coupling relevant for the Hall viscosity via:
\begin{align}
\mathbf{L}_i^{-q\nu;q\nu^\prime} \approx \frac{i}{4\hbar\omega_0} \left( 
\mathbf{B}_{i;++}^{q\nu;-q\nu^\prime}  
+\mathbf{B}_{i;+-}^{-q\nu;-q\nu^\prime}  
-\mathbf{B}_{i;-+}^{q\nu;q\nu^\prime}
-\mathbf{B}_{i;--}^{-q\nu;q\nu^\prime} 
\right)
\end{align}
In practice, this requires two ED calculations to evaluate, one with $q^\prime = q$, and one with $q^\prime = -q$. This follows from the fact that $\mathbf{B}_{i;++}^{q\nu;-q\nu^\prime}  = - \mathbf{B}_{i;--}^{-q\nu;q\nu^\prime} $ and $\mathbf{B}_{i;+-}^{-q\nu;-q\nu^\prime}  = -\mathbf{B}_{i;-+}^{q\nu;q\nu^\prime}$. Results of calculations of $\mathbf{L}$ are given in Fig.~\ref{fig:GLresults} in Supplementary Section S3.

\subsection{Supplemental Note 3. Numerical Results for L Couplings for Acoustic Phonons}

\begin{figure*}[b]
\includegraphics[width=\linewidth]{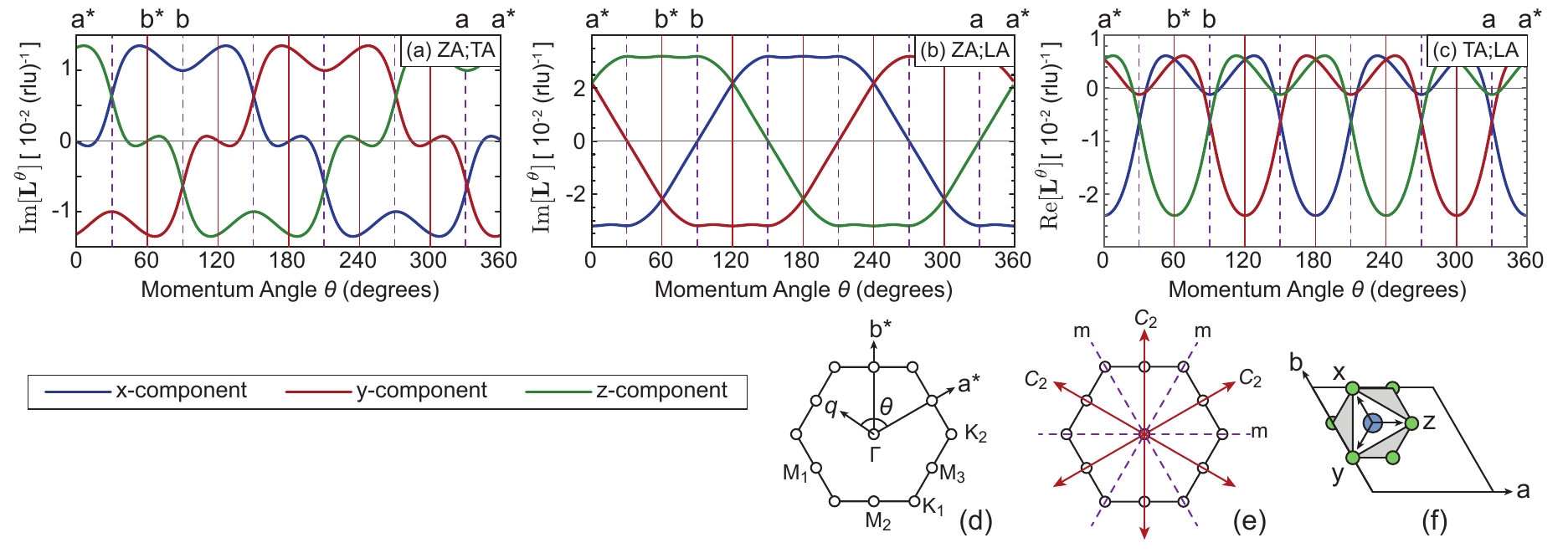}
\caption{{\bf Momentum angle dependence of the computed $\mathbf{L}_{i}^{\theta \nu;\nu^\prime}$ spin-phonon couplings.} (a)-(c): $(x,y,z)$ components of $\mathbf{L}_{i}^{\theta \nu;\nu^\prime}$. The pairs of acoustic modes $\nu;\nu^\prime$ = ZA, TA, and LA are labelled in the upper right of each panel. (d): first Brillouin zone showing the definition of $a^*$ and $b^*$ axes and $\theta$ (measured from the $a^*$ axis). (e): orientation, in $q$-space, of the relevant symmetry operations for the single-site couplings. (f) Single Ru site showing definition of cubic axes. 1 rlu = $4\pi/(\sqrt{3}a)$, reciprocal lattice unit.}
\label{fig:GLresults}
\end{figure*}

Here, we discuss the computed $q$-dependence of the $\mathbf{L}$ couplings for the acoustic modes. As noted in the main text, these couplings are sufficiently small in \rucl as to have negligible consequence on the thermal transport. As demonstrated in Supplementary Section S4, $\mathbf{L} \propto |q|$ for low momentum acoustic phonons. In the limit of small $q$, for the chosen gauge, the coupling is also equal at both Ru sites in the unit cell. Thus, we write:
\begin{align}
\mathbf{L}_i^{-q\nu;q\nu^\prime} = \mathbf{L}^{\theta}_{\nu;\nu^\prime} |q|
\end{align}
Fig.~\ref{fig:GLresults}(a-c) shows the computed momentum-dependence of $\mathbf{L}^{\theta}_{\nu;\nu^\prime}$. This quantity has units of inverse wavevector. The order of magnitude should be compared with the total angle-dependent reduced Hall viscosity $\overline{\eta}_\theta$ defined in Supplementary Section S5. The contribution to the latter quantity from $\mathcal{A}-\mathcal{A}$ bond-bond correlations 
 reaches several (rlu)$^{-1}$ at maximum. In contrast, the order of magnitude of the contribution from $\mathbf{L}$ interactions is $10^{-3}$ to $10^{-2}$ (rlu)$^{-1}$. For this reason, the two-phonon Raman interaction $\mathbf{L}$ does not make a significant contribution to the phonon thermal Hall effect. 
 
To investigate why these couplings are so small in \rucl, we performed calculations employing model electron-phonon couplings and idealized crystal fields, and confirmed that the largest contribution to $\mathbf{L}$ scales like $(\mathcal{H}_{\rm el-ph})^2/\lambda_{\rm Ru}^2$. The $\mathbf{L}$ couplings represent the nuclear Berry curvature associated with the evolution of the spin-orbital composition of the ground-state doublet at each site. The primary mechanism for this evolution is mixing of the $j_{1/2}$ states with $j_{3/2}$ states. The relatively large value of the SOC constant $\lambda_{\rm Ru}$ reduces this effect in \rucl. Thus, one can expect $\mathbf{L}$ to be suppressed in materials with strong SOC or otherwise energetically separated single-ion states.

The couplings depicted in Fig.~\ref{fig:GLresults} display the expected symmetries. $\mathbf{L}_{\rm ZA;TA}^\theta$ is even with respect to $C_2$ and odd with respect to $m$. $\mathbf{L}_{\rm ZA;LA}^\theta$ is odd with respect to $C_2$ and even with respect to $m$. $\mathbf{L}_{\rm TA;LA}^\theta$ is odd with respect to both $C_2$ and $m$. Within the chosen gauge, the leading contributions to $\mathbf{L}_{\rm ZA;TA}^\theta$ and $\mathbf{L}_{\rm ZA;LA}^\theta$ are imaginary, while $\mathbf{L}_{\rm TA;LA}^\theta$ is real. The combinations of the symmetries and gauge choice ensure that the $\mathbf{L}$ vectors wind around $q=0$. For each combination of bands, there is a component of $\mathbf{L}$ in the direction $\mathbf{u}_{q\nu} \times \mathbf{p}_{q\nu^\prime}$, which corresponds to a spin-phonon coupling like $\mathbf{S} \cdot (\mathbf{u} \times \mathbf{p})$ \cite{sheng2006theory}. For example, the primary atomic motion for the LA mode is in-plane, parallel to $q$. The primary motion for the TA mode is in-plane, perpendicular to $q$. There is a component of $\mathbf{L}_{\rm TA; LA}^\theta$ that points out of plane (along the cubic $\hat{x}+\hat{y}+\hat{z}$ direction).  However, we also find significant contributions to the $\theta$-dependence of $\mathbf{L}$ that depart from this form, which highlight the subtleties of spin-phonon coupling of spin-orbital moments.

\subsection{S4. Low-$q$ Scaling and Connection to Long-Wavelength Elastic Theory}
In this section, we discuss the low-$q$ scaling of various quantities for the acoustic phonons, and the relationship between the reduced Hall viscosity $\overline{\eta}_q$ and the long-wavelength viscosity tensor. In the long-wavelength limit, the Lagrangian density of an elastic medium is given by:
\begin{align}
\frac{\mathcal{L}}{V} = \frac{\rho}{2} \sum_{\alpha}  \dot{u}_\alpha - \frac{1}{2} \sum_{\alpha\beta\delta\gamma} C_{\alpha\beta\delta\gamma} \epsilon_{\alpha\beta}\epsilon_{\delta\gamma} + \sum_{\alpha\beta\delta\gamma} \eta_{\alpha\beta\delta\gamma} \epsilon_{\alpha\beta}\dot{\epsilon}_{\delta\gamma}
\label{eq:lagrangedens}
\end{align}
where $\dot{\epsilon} = \partial \epsilon / \partial t$ and $\rho$ is the mass density, $C_{\alpha\beta\delta\gamma}$ is the elasticity tensor, $\eta_{\alpha\beta\delta\gamma}$ is the viscosity tensor, and the strain field is:
\begin{align}
\epsilon_{\alpha\beta} = \frac{1}{2} \left(\frac{\partial u_\alpha}{\partial r_\beta} + \frac{\partial u_\beta}{\partial r_\alpha} \right)
\end{align}
This form of the Langrangian density can be recovered by taking the long-wavelength limit of equation (\ref{eq:hefffull}). For illustrative purpose, we neglect the $q$-dependence of the phonon eigenvectors by approximating $\partial e_{\alpha n}^{q\nu}/\partial q = 0$. This leads to the long-wavelength approximation:
\begin{align} \label{eq:eexpand}
e_{\alpha n}^{q\nu} \ e^{iq\cdot r_{\ell n}} \approx & \  \sqrt{\frac{m_n N }{\rho V}} \ \varepsilon_{\alpha\nu}^{q}\left[ \ 1  + i \ \vec{r}_{\ell n }  \cdot \vec{q} \  \right]+ O(q^2)
\end{align}
such that the displacement field is:
\begin{align}
\hat{u}_\alpha (\vec{r}) \approx & \  \frac{1}{\sqrt{N}} \sum_{q\nu} \hat{u}_{q\nu} \ \varepsilon_{\alpha \nu}^{q}\left[  1  + i  \  \vec{r}  \cdot \vec{q} \  \right]+ O(q^2)
\end{align}
where the phonon polarization coefficients $\varepsilon_{\alpha\nu}^{q}$ depend on the direction of the $q$-vector. We have utilized $\lim_{q\to 0} m_{q\nu} = \rho (V/N)$ for the acoustic modes, where $V/N$ is the volume of a unit cell.  The gauge choice described in Supplementary Section S2 corresponds to choosing $\varepsilon_{\alpha \nu}^{q}$ to be completely real for the ZA modes and completely imaginary for the TA and LA modes. In particular, for in-plane momenta:
\begin{align}
\varepsilon_{x,\rm LA}^{q} = i\frac{\vec{q}\cdot \hat{x}}{|\vec{q}|}  \ \ \ , \ \ \ \varepsilon_{y,\rm LA}^{q} = i\frac{\vec{q}\cdot \hat{y}}{|\vec{q}|}  \ \ \ , \ \ \varepsilon_{z,\rm LA}^{q} = 0
\\
\varepsilon_{x,\rm TA}^{q} = -i\frac{\vec{q}\cdot \hat{y}}{|\vec{q}|}  \ \ \ , \ \ \ \varepsilon_{y,\rm TA}^{q} = i\frac{\vec{q}\cdot \hat{x}}{|\vec{q}|}  \ \ \ , \ \ \varepsilon_{z,\rm TA}^{q} = 0
\\
\varepsilon_{x,\rm ZA}^{q} = 0  \ \ \ , \ \ \ \varepsilon_{y,\rm ZA}^{q} = 0 \ \ \ , \ \ \varepsilon_{z,\rm ZA}^{q} = 1
\end{align}
 In the limit $q\to 0$, the spin-phonon couplings for the acoustic modes must vanish, since those modes correspond to a uniform translation of the lattice. As a consequence, 
\begin{align}
    \lim_{q\to 0}\mathbb{A}_{ij}^{qv} = & \  \sum_{\alpha \ell n} \sqrt{\frac{\hbar N}{2\rho V\omega_{qv}}}\ \mathbb{A}_{ij}^{\alpha\ell n} \ \varepsilon_{\alpha \nu}^{q}  = 0
\end{align}
which leads to:
\begin{align}
\mathbb{A}_{ij}^{qv} \approx i \sum_{\alpha \ell n} \sqrt{\frac{\hbar N}{2\rho V\omega_{qv}}}\ \mathbb{A}_{ij}^{\alpha\ell n} \left[ \ \varepsilon_{\alpha\nu}^{q} \left(  \vec{r}_{\ell n} \cdot \vec{q}\right)  \ \right] \equiv \mathbb{A}_{ij}^{\theta \nu} \ |q|^{1/2}
\end{align}
Here, we have assumed that $\omega_q \propto |q|$. This is satisfied for all acoustic modes at sufficiently small $|q|$. As a consequence, $\mathbb{A}_{ij}^{qv}  \propto |q|^{1/2}$. A similar argument leads to the following scaling of the couplings and operators:
\begin{align}
\mathbb{A} \propto & \ |q|^{1/2}
\\
\mathcal{A} \propto & \ |q|^{1/2}
\\
\mathbf{L} \propto & \ |q|
\\
\mathcal{L} \propto & \ |q|^2
\\
\overline{\eta} \propto & \ |q|
\end{align}
The extra factor of $q$ in the $\mathcal{L}$ operators compared to the $\mathbf{L}$ couplings arises from the factor of $\omega_q$ in equation (\ref{eqn:Ldefsup}). We have verified that this scaling is satisfied in our numerical results.

In the same long-wavelength limit, the strain field is:
\begin{align}
\epsilon_{\alpha\beta} = \frac{i}{2} \frac{1}{\sqrt{N}}\sum_{q\nu} u_{q\nu} \left( \varepsilon_{\alpha\nu}^{q } q_\beta + \varepsilon_{\beta\nu}^{q} q_\alpha\right)
\end{align}
By comparing equations (\ref{eq:lagrange}) and (\ref{eq:lagrangedens}), we identify:
\begin{align}
\overline{\eta}_q^{\nu\nu^\prime} = \sum_{\alpha\beta\delta\gamma} \frac{\left( \varepsilon_{\alpha\nu}^{-q } q_\beta + \varepsilon_{\beta\nu}^{-q} q_\alpha\right)  \left( \varepsilon_{\delta\nu^\prime}^{q } q_\gamma + \varepsilon_{\gamma\nu^\prime}^{q} q_\delta\right) }{8\rho \sqrt{\omega_{q\nu}\omega_{q\nu^\prime}}} \left( \eta_{\alpha\beta\delta\gamma} - \eta_{\delta\gamma\alpha\beta} \right)
\end{align}
It it then instructive to consider the approximate angular dependence of the reduced Hall viscosity. We define:
\begin{align}
\eta_{\alpha\beta\delta\gamma}^H = \frac{1}{8}\left( 
\eta_{\alpha\beta\delta\gamma}
+\eta_{\alpha\beta\gamma\delta}
+\eta_{\beta\alpha\delta\gamma}
+\eta_{\beta\alpha\gamma\delta}
-\eta_{\delta\gamma\alpha\beta}
-\eta_{\gamma\delta\alpha\beta}
-\eta_{\delta\gamma\beta\alpha}
-\eta_{\gamma\delta\beta\alpha}
\right)
\end{align}
Then, for $q_x \ || \ a$, and $q_y\  || \ b^*$:
\begin{align}
\overline{\eta}_q^{\rm ZA; TA} = & \ \frac{i}{\rho \sqrt{c_{\rm ZA}c_{\rm TA}}|q|^2}\left[ 
\eta_{xzxy}^Hq_x^3
+(\eta_{xxxz}^H+\eta_{xzyy}^H+\eta_{yzxy}^H)q_x^2 q_y
+ (\eta_{xyxz}^H+\eta_{xxyz}^H+\eta_{yzyy}^H)q_xq_y^2
+\eta_{xyyz}^Hq_y^3
\right]
\\
\overline{\eta}_q^{\rm ZA; LA} = & \ \frac{i}{\rho \sqrt{c_{\rm ZA}c_{\rm LA}}|q|^2}\left[ 
\eta_{xzxx}^H q_x^3
+ \left( 2\eta_{xzxy}^H  +\eta_{yzxx}^H \right) q_x^2 q_y 
+ \left(  2\eta_{yzxy}^H +\eta_{xzyy}^H \right)q_xq_y^2
+ \eta_{yzyy}^H q_y^3
\right]
\\
\overline{\eta}_q^{\rm TA; LA} = & \ \frac{1}{\rho \sqrt{c_{\rm TA}c_{\rm LA}}|q|}\left[ 
\eta_{xyxx}^H  \ q_x^2
+\eta_{yyxx}^H  \ q_xq_y 
+\eta_{yyxy}^H \ q_y^2 
\right]
\end{align}
where $c_\nu = \partial \omega_{q\nu}/(\partial q)$ is the speed of sound for each band. From this, we see that the reduced Hall viscosities are particular angle-dependent linear combinations of the long-wavelength Hall viscosities. As discussed in Supplementary Section S5, the numerical estimates of $\overline{\eta}$ follow approximately these forms, but additional momentum-dependence of the phonon eigenvectors $e_{\alpha n}^{q\nu}$ not included equation (\ref{eq:eexpand}) somewhat enriches the angular-dependence of $\overline{\eta}$.

\subsection{Supplemental Note 5. Evaluation of Acoustic Phonon Thermal Conductivities}
{\it Effective Hamiltonian:} At sufficiently low temperatures, only the acoustic phonon bands contribute significantly to the phonon thermal transport. Focusing on these bands, the effective phonon Hamiltonian may be written:
\begin{align}
    \mathcal{H}_{\rm eff} = & \ \frac{1}{2}\sum_q
    \begin{pmatrix}
        \mathbf{a}_q^\dagger & \mathbf{a}_{-q}
    \end{pmatrix}
\mathcal{H}_q
    \begin{pmatrix}
        \mathbf{a}_q \\ \mathbf{a}_{-q}^\dagger
    \end{pmatrix}
    \\
    \mathcal{H}_q =& \     \begin{pmatrix}
        \Lambda_{q} +\frac{i}{2}\left( \Lambda_{q} \mathbb{N}_q +\mathbb{N}_q \Lambda_{q}\right) 
        & 
        \frac{i}{2}\left( \Lambda_{q}  \mathbb{N}_q - \mathbb{N}_q\Lambda_{q} \right)
        \\
        -\frac{i}{2}\left( \Lambda_{q} \mathbb{N}_q- \mathbb{N}_q \Lambda_{q} \right)
        &
        \Lambda_{q} -\frac{i}{2}\left( \Lambda_{q} \mathbb{N}_q+ \mathbb{N}_q \Lambda_{q}\right) 
    \end{pmatrix}
\end{align}
where $\mathbf{a}_q$ is a column vector of $a_{q\nu}$ operators for different phonon bands $\nu$, and $\Lambda_q$ is a diagonal matrix of phonon energies:
\begin{align}
\Lambda_q = \hbar \begin{pmatrix}
 \omega_{q,\rm ZA} & 0 & 0
\\
0 & \omega_{q,\rm TA} & 0
\\
0 & 0 & \omega_{q,\rm LA}
\end{pmatrix} = \hbar \begin{pmatrix}
c_{\rm ZA} & 0 & 0
\\
0 & c_{\rm TA} & 0
\\
0 & 0 & c_{\rm LA}
\end{pmatrix} |q|
\end{align}
Here, we have defined $\hbar\omega_{q\nu} = \hbar c_\nu|q|$, where $c_\nu = \partial \omega_{q\nu} / \partial|q|$ is the speed of sound for each band, which we estimate as $\hbar c_\mathrm{ZA} = 6.6$\,meV/rlu, $\hbar c_\mathrm{TA} = 11.7$\,meV/rlu, and $\hbar c_\mathrm{LA} = 28.2$\,meV/rlu from the {\it ab-initio} phonon calculations described in Supplementary Section S2. We have ignored the symmetric part of the phonon self-energy in the adiabatic limit because it only results in a very weak renormalization of the phonon velocities. The reduced Hall viscosity matrix is:
\begin{align}
\mathbb{N}_q = \begin{pmatrix}
0 & \overline{\eta}_q^{\rm ZA;TA} & \overline{\eta}_q^{\rm ZA;LA}
\\
\overline{\eta}_q^{\rm TA;ZA} & 0 & \overline{\eta}_q^{\rm TA;LA}
\\
\overline{\eta}_q^{\rm LA;ZA} & \overline{\eta}_q^{\rm LA;TA} & 0
\end{pmatrix} 
= \begin{pmatrix}
0 & \overline{\eta}_\theta^{\rm ZA;TA} & \overline{\eta}_\theta^{\rm ZA;LA}
\\
\overline{\eta}_\theta^{\rm TA;ZA} & 0 & \overline{\eta}_\theta^{\rm TA;LA}
\\
\overline{\eta}_\theta^{\rm LA;ZA} & \overline{\eta}_\theta^{\rm LA;TA} & 0
\end{pmatrix} |q|
\end{align}
where:
\begin{align}
\overline{\eta}_q^{\nu \nu^\prime} (T)= & \ \sum_n \frac{e^{-\beta E_n}}{\mathcal{Z}}\left[\frac{1}{N} \sum_i \langle n |  \mathbf{L}_i^{-q\nu;q\nu^\prime} \cdot \mathbf{S}_i |n \rangle\right]
\nonumber \\
& \ + \frac{i}{2} \sum_{nm} \frac{e^{-\beta E_n}-e^{-\beta E_m} }{\mathcal{Z}} \left[  
        \frac{\langle n|  \mathcal{A}_{q\nu}^\dagger |m\rangle \langle m | \mathcal{A}_{q\nu^\prime}|n \rangle-\langle n|  \mathcal{A}_{q\nu^\prime} |m\rangle\langle m| \mathcal{A}_{q\nu}^\dagger |n \rangle}{(E_n-E_m)^2}\right]       
\end{align}
and $\mathcal{A}$ are the linear spin-phonon coupling operators and $\mathbf{L}$ are the vectors defining the two-phonon Raman interaction. $|n\rangle$ and $|m\rangle$ are unperturbed spin states, and $\mathcal{Z}$ is the unperturbed spin partition function. As described in Supplementary Section S4, $\overline{\eta}_q \propto |q|$; we therefore define $\overline{\eta}_q = \overline{\eta}_\theta |q|$, where $\overline{\eta}_\theta$ captures the dependence on in-plane momentum direction. In the chosen gauge, $\overline{\eta}_\theta^{\rm ZA;TA} =  \overline{\eta}_\theta^{\rm TA;ZA}$ and $ \overline{\eta}_\theta^{\rm ZA;LA} =\overline{\eta}_\theta^{\rm LA;ZA}$ are purely imaginary, while $ \overline{\eta}_\theta^{\rm TA;LA} = -\overline{\eta}_\theta^{\rm LA;TA}$ is purely real.  The approximate form of the reduced Hall viscosity in terms of the long-wavelength viscosity tensor is described in Supplementary Section S4. We have defined the spin-phonon coupling operators $\mathcal{A}$ and $\mathcal{L}$ to have units of energy for convenience, which leads to $\overline{\eta}_q$ being unitless, while $\overline{\eta}_\theta$ has units of inverse wavevector. 

{\it Comparison with long-wavelength limit:} Returning to the long-wavelength approximation in Supplementary Section S4, the approximate form of the reduced Hall viscosities for different field directions can be deduced from symmetry \cite{birss1962property}. In the high-field polarized phase for a magnetic field $\mathbf{B} \ || \ a$, the magnetic point group is $2^\prime/m^\prime$, which leads to:
\begin{align}
\overline{\eta}_\theta^{\rm ZA; TA} = & \ \frac{i}{\rho \sqrt{c_{\rm ZA}c_{\rm TA}}|q|^3}\left[ 
 \eta_{xzxy}^Hq_x^3 
- (\eta_{yyyz}^H+\eta_{xzxy}^H+\eta_{yzxx}^H)q_x q_y^2 
\right]
\\
\overline{\eta}_\theta^{\rm ZA; LA} = & \ \frac{i}{\rho \sqrt{c_{\rm ZA}c_{\rm LA}}|q|^3}\left[ 
 \left( 2\eta_{xzxy}^H  +\eta_{yzxx}^H \right) q_x^2 q_y  
- \eta_{yyyz}^H q_y^3 
\right]
\\
\overline{\eta}_\theta^{\rm TA; LA} = & \ \frac{1}{\rho \sqrt{c_{\rm TA}c_{\rm LA}}|q|^2}\left[ 
\eta_{xyxx}^H  \ q_x^2 
+\eta_{yyxy}^H \ q_y^2  
\right]
\end{align}
where $q_x \ ||\ a$, and $q_y \ || \ b^*$. This allows a finite $\kappa_{xy}$. In contrast, in the zero-field zigzag phase or high-field polarized phase with magnetic field $\mathbf{B} \ || \ b^*$, the magnetic point group is $2/m$, which allows:
\begin{align}
\overline{\eta}_\theta^{\rm ZA; TA} = & \ \frac{i}{\rho \sqrt{c_{\rm ZA}c_{\rm TA}}|q|^3}\left[ 
(\eta_{xxxz}^H+\eta_{xzyy}^H+\eta_{yzxy}^H)q_x^2 q_y 
-\eta_{yzxy}^Hq_y^3 
\right]
\\
\overline{\eta}_\theta^{\rm ZA; LA} = & \ \frac{i}{\rho \sqrt{c_{\rm ZA}c_{\rm LA}}|q|^3}\left[ 
\left(  2\eta_{yzxy}^H +\eta_{xzyy}^H \right)q_xq_y^2 
- \eta_{xxxz}^H q_x^3 
\right]
\\
\overline{\eta}_\theta^{\rm TA; LA} = & \ \frac{1}{\rho \sqrt{c_{\rm TA}c_{\rm LA}}|q|^2}\left[ 
\eta_{yyxx}^H  \ q_xq_y  
\right]
\end{align}
In our numerical results, we find that $\overline{\eta}_\theta^{\nu;\nu^\prime}$ follow these forms approximately, but also have additional higher order $\sin 4\theta$, $\cos 4\theta$, $\sin 6\theta$, $\cos 6\theta$, etc.~contributions resulting from the momentum-dependence of the phonon eigenvectors. Nonetheless, we may estimate the long-wavelength Hall viscosity tensor components by fitting the computed $\theta$-dependence of  $\overline{\eta}_\theta^{\nu;\nu^\prime}$. The results are shown in Fig.~\ref{fig:eta_deconstruct}. As expected, all components of $\eta_{\alpha\beta\delta\gamma}^H$ vanishes at zero field. For $\mathbf{B} \ || \ a$, all components of the Hall viscosity tensor are enhanced for $B > B_c$, reaching a maximum magnitude in the range $10 - 12$ T. This corresponds to the maximum in $\kappa_{xy}(B)$. For $\mathbf{B} \ || \ b^*$, instead several components of the Hall viscosity tensor change sign at $B_c$, displaying maximum magnitudes both above and below $B_c$. Although $\kappa_{xy} = 0 $ for this field direction, the Hall viscosity is finite.


\begin{figure}[t]
\includegraphics[width=\linewidth]{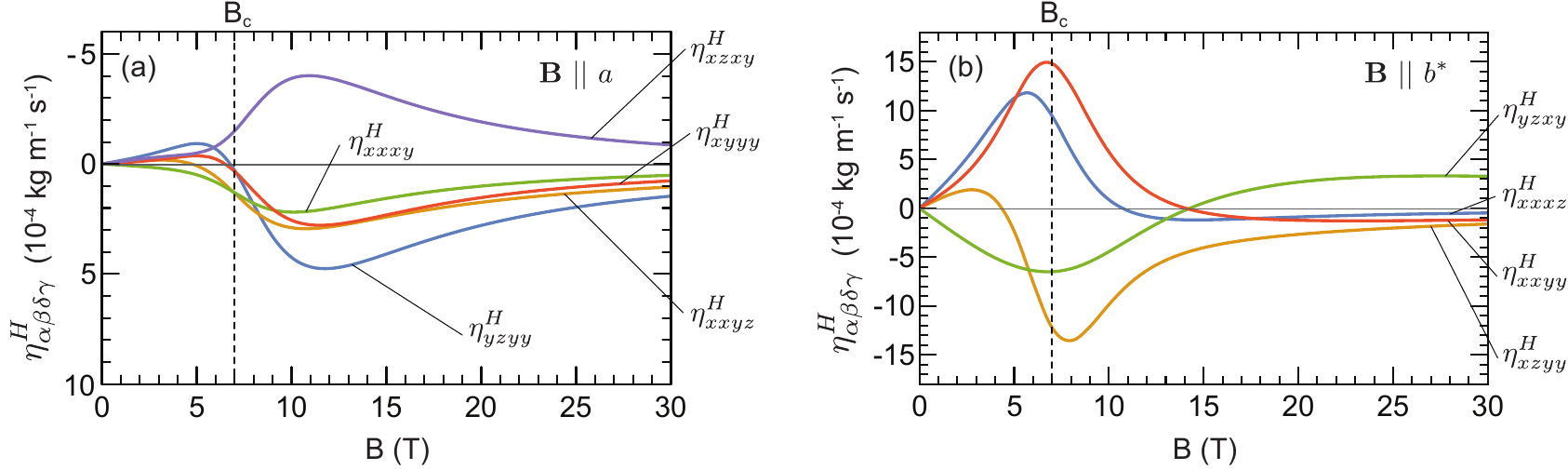}
\caption{ {\bf Evolution of different components of the long-wavelength Hall viscosity tensor.} Results obtained by fitting computed $\overline{\eta}_\theta^{\nu;\nu^\prime}$ to approximate long-wavelength forms for (a): $\mathbf{B} \ || \ a$ and (b): $\mathbf{B} \ || \ b^*$.  Results based on zero-temperature bond correlations computed using exact diagonalization of the spin model discussed in the main text.}
\label{fig:eta_deconstruct}
\end{figure}


{\it Evaluation of $\kappa_{xy}$:} Following \cite{matsumoto2014thermal,kondo2020non,toth2015linear}, we define a matrix $\mathbf{g}$, which is:
\begin{align}
    \mathbf{g} = \left[\begin{pmatrix}
        \mathbf{a}_q \\ \mathbf{a}_{-q}^\dagger
    \end{pmatrix},\begin{pmatrix}
        \mathbf{a}_q^\dagger & \mathbf{a}_{-q}
    \end{pmatrix} \right] = 
    \begin{pmatrix}
        \mathbb{I}_{N\times N} & 0 \\ 0 & -\mathbb{I}_{N\times N}
    \end{pmatrix}
\end{align}
where $\mathbb{I}_{N\times N}$ is the identity matrix of dimension $N = 3$, which is the number of phonon bands. The phonon eigenvectors $\xi_{q\nu}^{L}$ and $\xi_{q\nu}^{R}$ are the left and right eigenvectors of $\mathbf{g}\mathcal{H}_q$ with eigenvalues $\lambda_{q\nu} \approx \pm \omega_{q\nu}$, and the 2D phonon Berry curvature is given by:
\begin{align}\label{eq:BCsup}
    \Omega_{q\nu}^{\rm BC} = -\text{sign}(\lambda_{q\nu}) \ \text{Im}\left[\left(\frac{\partial}{\partial q_x}\xi_{q\nu}^ {L}\right)\mathbf{g}\left(\frac{\partial}{\partial q_y}\xi_{q\nu}^{R}\right) -  \left(\frac{\partial}{\partial q_y}\xi_{q\nu}^{L}\right)\mathbf{g}\left(\frac{\partial}{\partial q_x}\xi_{q\nu}^{R}\right)\right] \equiv \Omega_{\nu}^\theta |q|
\end{align}
This is the Berry curvature of the phonon bands associated with a $q$-dependent remixing of the phonons induced by the Hall viscosity terms in $\mathcal{H}_q$ [not to be confused with the nuclear Berry curvature defined in equation (\ref{eq:nuclearBC}), which determines the Hall viscosity]. As discussed in more detail below, we find $\Omega_{q\nu}^{\rm BC} \propto |q|$. Thus, following \cite{li2023magnons}:
\begin{align}
\frac{\kappa_{xy}^{2\mathrm{D}}}{T} = & \ -\frac{3\kappa_Q}{2\pi^3}  \sum_\nu \int_\mathrm{BZ(2D)} d\vec{q}\  c_2\left[ n_\mathrm{B}(\hbar\omega_{q\nu})\right] \Omega_{q\nu}^{\rm BC}
\end{align}
where:
\begin{align}
c_2(x) =& \  (1+x) \left[\text{ln} \left( \frac{1+x}{x}\right) \right]^2 - \left[\text{ln}(x)\right]^2 - 2\text{Li}_2(-x)
\\
n_\mathrm{B}(E) =& \  \frac{1}{e^{\frac{E}{k_\mathrm{B}T}} -1 }
\end{align}
where $\kappa_Q = \pi k_\mathrm{B}^2/(6\hbar)$ is the quantum of thermal conductivity and Li$_2$ is the polylog function. For low temperatures, the acoustic phonons at the edge of the Brillouin contribute little to the integral, because they are sufficiently high in energy. As such, the integral can be extended to infinite $|q|$, and written:
\begin{align}
\frac{\kappa_{xy}^{2\mathrm{D}}}{T} \approx &\  -\frac{3\kappa_Q }{2\pi^3}  \sum_\nu \int_0^{2\pi}  \Omega_\nu^\theta \ d\theta \int_0^{\infty} |q|^2 \ c_2(f(\hbar c_\nu |q| )) \  d|q|
\end{align}
using $\Omega_{q\nu}^{\rm BC} = \Omega_\nu^\theta |q|$ and $\hbar \omega_{q\nu} = \hbar c_\nu|q|$. Introducing $x =  \frac{\hbar c_\nu|q|}{k_\mathrm{B}T}$ gives:
\begin{align}
\frac{\kappa_{xy}^{2\mathrm{D}}}{T} 
\approx &\  -\frac{3\kappa_Q }{2\pi^3}  \sum_\nu \left( \frac{k_\mathrm{B}T}{\hbar c_\nu} \right)^3\int_0^{2\pi}  \Omega_\nu^\theta \ d\theta \int_0^{\infty} x^2 \ c_2 \left[ \frac{1}{e^x-1}\right] \  dx \ 
\\
\approx &\ - 2.00655  \ \kappa_Q  \sum_\nu \left( \frac{k_\mathrm{B}T}{\hbar c_\nu} \right)^3\int_0^{2\pi}  \Omega_\nu^\theta \ d\theta
\label{eq:suppkxy}
\end{align}

{\it Conditions for finite $\kappa_{xy}$:} It is finally instructive to consider the restrictions on $\overline{\eta}_q$ that allow for a finite $\kappa_{xy}$. To do so, we consider the case where the three phonon bands are non-degenerate, and treat the effects of the Hall viscosity perturbatively. In particular,
\begin{align}
    \mathcal{H}_q^{(0)} =   \begin{pmatrix}
        \Lambda_{q}
        & 
       0
        \\
        0
        &
        \Lambda_{q} \end{pmatrix} \ \ \ \ , \ \ \ \   \mathcal{H}_q^{(1)} = \begin{pmatrix}
       \frac{i}{2}\left( \Lambda_{q} \mathbb{N}_q +\mathbb{N}_q \Lambda_{q}\right) 
        & 
        \frac{i}{2}\left( \Lambda_{q}  \mathbb{N}_q - \mathbb{N}_q\Lambda_{q} \right)
        \\
        -\frac{i}{2}\left( \Lambda_{q} \mathbb{N}_q- \mathbb{N}_q \Lambda_{q} \right)
        &
         -\frac{i}{2}\left( \Lambda_{q} \mathbb{N}_q+ \mathbb{N}_q \Lambda_{q}\right) 
    \end{pmatrix}
\end{align}
We then consider the perturbative expansion of the eigenvectors of $\mathbf{g} \mathcal{H}_q$:
\begin{align} \label{eq:perturbation}
\xi_{q\nu}^{R}\approx \xi_{q\nu}^{R,0} + \sum_{\nu^\prime \neq \nu} \xi_{q\nu^\prime}^{R,0}  \frac{\xi_{q\nu^\prime}^{L,0} \mathcal{H}_q^{(1)} \xi_{q\nu}^{R,0} }{\lambda_{q\nu}^0 - \lambda_{q\nu^\prime}^0} 
+ \sum_{\nu^{\prime\prime},\nu^\prime \neq \nu} \xi_{q\nu^{\prime}}^{R,0}  \frac{\xi_{q\nu^{\prime}}^{L,0} \mathcal{H}_q^{(1)} \xi_{q\nu^{\prime\prime}}^{R,0} \xi_{q\nu^{\prime\prime}}^{L,0} \mathcal{H}_q^{(1)} \xi_{q\nu}^{R,0} }{(\lambda_{q\nu}^0 - \lambda_{q\nu^{\prime\prime}}^0)(\lambda_{q\nu}^0 - \lambda_{q\nu^\prime}^0)} + ...
\end{align}
and finally compute the phonon Berry curvature $\Omega_{q\nu}^{\rm BC}$ using equation (\ref{eq:BCsup}). We find the leading contribution to scale linearly with $|q|$, such that $\Omega_{q\nu}^{\rm BC} =  \Omega_\nu^\theta |q| $. The latter quantity is:
\begin{align}
\Omega_\nu^\theta = & \ \left[f_{1\nu}  \ \overline{\eta}_\theta^{\rm ZA;TA}  \overline{\eta}_\theta^{\rm TA;LA}\frac{\partial  \overline{\eta}_\theta^{\rm ZA;LA} }{\partial \theta} 
+f_{2\nu}  \ \overline{\eta}_\theta^{\rm ZA;TA} \frac{\partial  \overline{\eta}_\theta^{\rm TA;LA} }{\partial \theta}   \overline{\eta}_\theta^{\rm ZA;LA}
+f_{3\nu}  \ \frac{\partial  \overline{\eta}_\theta^{\rm ZA;TA} }{\partial \theta}   \overline{\eta}_\theta^{\rm TA;LA}  \overline{\eta}_\theta^{\rm ZA;LA}\right]
\label{eq:BCang}
\end{align}
where $f_{1\nu}, f_{2\nu}$, and $f_{3\nu}$ are real band-dependent functions that depend only on the relative sound velocities of the bands (not on the $\theta$ or $q$). There are several important observations underlying this expression. First, as noted above, $\Omega_{q\nu}^{\rm BC} \propto |q|$. This agrees with our numerical, which confirms the perturbative regime is appropriate. Second, we see that a finite $\Omega_{q\nu}^{\rm BC}$ requires all three elements of the Hall viscosity matrix $\mathbb{N}_q$ to be finite. That is, the coupling of any two bands is not sufficient to induce a finite Berry curvature in any of the bands in the 2D limit. This is true because a smooth gauge exists at low $q$ (except at $q=0$) such that all $\overline{\eta}_q$ are either completely real or completely imaginary. At first order in $\mathcal{H}_q^{(1)}$, the complex phases of the phonon eigenvectors therefore do not wind as a function of $\theta$.  Instead, the Berry curvature arises from the combination of the first and second order terms in equation (\ref{eq:perturbation}). For example, the ZA phonon band may acquire a finite mixture of the TA mode due to the interband Hall viscosity matrix elements $i\ \mathbb{N}_q$. The first order contribution to the mixing corresponds to $i\overline{\eta}_\theta^{\rm ZA;TA}$, which is real. The second order contribution corresponds to $(i\overline{\eta}_\theta^{\rm ZA;LA})(i\overline{\eta}_\theta^{\rm LA;TA})$, which is imaginary. Thus the phase of the complex coefficient representing the small mixing of the TA mode into the ZA eigenvector can wind around the Brillouin zone provided the first and second order terms have complementary momentum-angle dependence.

As can be seen above, $\kappa_{xy}^{\rm ph} \propto \int \Omega_\nu^\theta \propto \overline{\eta}^3$, which scales roughly as $ \overline{\eta}^3 \sim \sum_n ( \langle n| \mathcal{A}|g\rangle/\Delta_n)^6 $, where $\Delta_n$ is the energy of the $n$th excited state of the spin system. For this reason, accurate estimates of the magnitude of $\kappa_{xy}^{\rm ph}$ require precise modeling of the evolution of the spin excitation energies.


\begin{figure}[t]
\includegraphics[width=\linewidth]{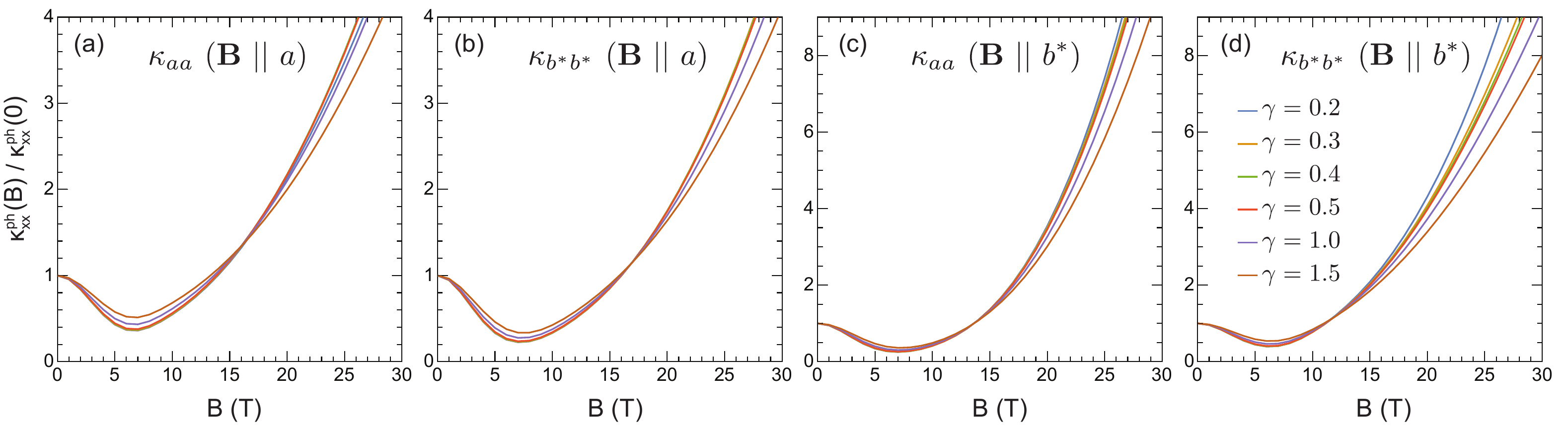}
\caption{ {\bf Comparison of computed longitudinal phonon thermal conductivity as a function of Lorentzian broadening.} Results obtained by applying broadening ($\gamma$ in meV) to the poles of the dynamical bond correlation functions obtained from exact diagonalization, and extrapolating to zero frequency, as discussed in the text. $\kappa_{xx}^{\rm ph} (B) /\kappa_{xx}(0)$ is relatively insensitive to choice of broadening. $\gamma = 0.5$\,meV is employed in the main text. }
\label{fig:broadening}
\end{figure}


{\it Evaluation of $\kappa_{xx}$:} In the 2D limit, the longitudinal conductivity is given by:
\begin{align}
\kappa_{\alpha\alpha}^{\rm 2D} =& \  \sum_\nu \int_{\rm BZ(2D)} d\vec{q} \ \frac{\hbar^2 \omega_{q\nu}^2}{k_BT^2}\frac{e^{\frac{\hbar\omega_{q\nu}}{k_BT}}}{\left( e^{\frac{\hbar\omega_{q\nu}}{k_BT}}-1\right)^2} (\nabla_q \omega_{q\nu} \cdot \hat\alpha)^2\tau_{q\nu} 
\end{align}
where $\tau_{q\nu}$ is the phonon lifetime. Introducing $x =  \frac{\hbar c_\nu|q|}{k_\mathrm{B}T}$:
\begin{align}
\kappa_{\alpha\alpha}^{\rm 2D} = & \  \frac{1}{(2\pi)^2}\frac{k_B^2 T}{\hbar^2}\sum_\nu \int_0^\infty dx  \frac{x^2 e^{x}}{\left( e^{x}-1\right)^2}  \int_0^{2\pi} d\theta (\hat{q} \cdot \hat\alpha)^2 \left(\hbar c_\nu |q|\tau_{q\nu} \right)
\\
= & \ \frac{k_B^2 T}{12 \hbar^2}\sum_\nu \int_0^{2\pi} d\theta \ (\hat{q} \cdot \hat\alpha)^2 \ \tau_\nu^\theta
\end{align}
where $\tau_\nu^\theta \equiv \hbar c_\nu |q|\tau_{q\nu}$ is the angle-dependent relative phonon lifetime. It is approximately given by the imaginary part of the diagonal components of the phonon self-energy evaluated at the bare phonon frequency:
\begin{align}
\tau_\nu^\theta = \frac{\hbar c_\nu |q|}{\text{Im}[\Pi_{q}^{\nu\nu}(\omega_{q\nu})]}
\end{align}
which is independent of $|q|$. Here, we assume that spin-phonon scattering is the most significant mechanism affecting the thermal transport lifetime at low temperatures. Additional contributions to the scattering rate from impurities and phonon anharmonicity may be relevant for quantitative calculations. Finally, we comment on the numerical evaluation of $\text{Im}[\Pi_{q}^{\nu\nu}(\omega_{q\nu})]$ using exact diagonalization (ED) of the spin model discussed in the main text. From ED, we obtain $\Pi_{q}^{\nu\nu}(\omega_{q\nu})$ as discrete poles, which are Lorentzian broadened to extrapolate to low frequency:
\begin{align}
\text{Im}[\Pi_{q}^{\nu\nu}(\omega)] \approx \sum_n \frac{1}{\pi}\frac{\gamma A_n}{(E_n-\omega)^2+\gamma^2}
\end{align}
where $\gamma$ is the broadening, and $A_n$ is the amplitude of the correlation function for excited state $|n\rangle$. The discrete poles contributing to this function remain above $E_n$ = 1\,meV even at $B_c$ due to finite size effects, which is above the energies of the phonons with appreciable  population at low temperatures. The broadening results in an approximately linear dependence of the low-frequency $\text{Im}[\Pi_{q}^{\nu\nu}(\omega_{q\nu})]$ on $\gamma$. As a consequence, absolute magnitudes of the spin-phonon scattering rate cannot be resolved, but relative values remain robust. This is demonstrated in Fig.~\ref{fig:broadening}, where we plot $\kappa_{xx}^{\rm ph} (B) /\kappa_{xx}^{\rm ph}(0)$ derived from different choices of broadening. The qualitative field dependence is preserved over a wide range of choices of $\gamma$.

\end{widetext}

\end{document}